\documentclass[A4paper,11pt]{article}
\usepackage{graphics,graphicx,epsfig,wrapfig}
\usepackage{longtable}  
\usepackage{amsmath,verbatim,amssymb,color,lscape}
\usepackage{kotex}
\usepackage{natbib}
\usepackage{lastpage}
\usepackage{multirow} 
\usepackage{authblk}
\usepackage{bm}
\usepackage{subfig}
\usepackage{supertabular}
\usepackage[ruled,vlined]{algorithm2e}
\usepackage[normalem]{ulem}
\usepackage[export]{adjustbox}
\usepackage{float}
\usepackage{makecell}
\usepackage{url}

\newcommand{\hide}[1]{}

\makeatletter

\usepackage[outerbars,color]{changebar}
\ifx\pdfoutput\undefined
\else\ifnum\pdfoutput>0
  \usepackage{pdfcolmk}
\fi\fi
\cbcolor{black}

\usepackage[margin=1.0in]{geometry}
\usepackage{tikz}
\usetikzlibrary{bayesnet}
\usetikzlibrary{fit,positioning}

\newfont{\rmm}{cmr10 at 11pt}
\rmm

\title{Hierarchical Latent Space Item Response Model for Analyzing Mental Health Vulnerability of Elementary School Students in South Korea}

\author[1]{Soyeon Park}
\author[1]{Seoyoung Shin}
\author[3]{Minjeong Jeon}
\author[4]{Hyoun Kyoung Kim}
\author[1,2]{Ick Hoon Jin\footnote{e-mail: ijin@yonsei.ac.kr. Corresponding author}}

\affil[1]{Department of Statistics and Data Science, Yonsei University. Seoul. South Korea.}
\affil[2]{Department of Applied Statistics, Yonsei University. Seoul. South Korea.}
\affil[3]{School of Education and Information Studies, University of California, Los Angeles. USA.}
\affil[4]{Department of Child and Family Studies, Yonsei University. Seoul. South Korea.}
\date{}

\begin{document}

\begin{titlepage}
\centering

\vspace*{2cm}

{\LARGE \textbf{Hierarchical Latent Space Item Response Model for Analyzing Mental Health Vulnerability of Elementary School Students in South Korea}}

\vspace{2cm}

Soyeon Park$^1$, Seoyoung Shin$^1$, Minjeong Jeon$^3$, Hyoun Kyoung Kim$^4$, and Ick Hoon Jin$^{1,2,*}$

\vspace{1.5cm}

$^1$Department of Statistics and Data Science, Yonsei University, Seoul, South Korea

$^2$Department of Applied Statistics, Yonsei University, Seoul, South Korea

$^3$School of Education and Information Studies, University of California, Los Angeles, USA

$^4$Department of Child and Family Studies, Yonsei University, Seoul, South Korea

\vspace{2cm}

$^*$Corresponding Author:\\
Ick Hoon Jin\\
Department of Applied Statistics, \\
Department of Statistics and Data Science,\\
Yonsei University.\\
50 Yonsei-ro, Seodaemun-gu, Seoul 03722, South Korea\\
e-mail: ijin@yonsei.ac.kr

\vspace{2cm}

\textbf{Short Running Head:} Hierarchical LSIRM for Mental Health of Elementary Students

\end{titlepage}

\maketitle

% @misc{}

\begin{abstract}
Mental health difficulties among elementary school students represent a growing public health concern in South Korea, yet analytical tools for identifying school-specific vulnerability patterns from item response data remain limited. We propose the hierarchical latent space item response model (HLSIRM), which adds hierarchical respondent effects and an inner-product latent interaction for signed respondent–item associations, yielding a unified interaction map that separates school/individual main effects from school/individual–item interactions. We apply HLSIRM to mental health vulnerability data from 2,210 elementary school students across 35 schools in Incheon, South Korea. Clustering item vectors by directional similarity identifies four empirically derived vulnerability domains. School-level analysis reveals that the absence of counseling experience is the primary vulnerability domain aligned with most school vectors, while stress, depression, and smartphone dependency concentrate in specific schools. Within-school analysis demonstrates how individual student positions in the interaction map translate into targeted intervention strategies that address school-specific needs.
\end{abstract}

\newpage

\section{Introduction}\label{sec:introduction}

% % === OPENING: Establish the Problem and Research Gap ===

Mental health difficulties in childhood and adolescence constitute a major public health concern, as excessive stress and psychological distress during these developmental periods are strongly associated with elevated suicide risk \citep{san2019association, o2020stress}. In South Korea, this concern is particularly acute: suicide has remained the leading cause of death among adolescents aged 10--19 since 2007, accounting for 60.2\% of all deaths in this age group in 2024 \citep{stat_2024_death}. Unlike previously prominent causes of mortality such as diseases or traffic accidents, which were substantially reduced through systemic and infrastructural interventions, mental health challenges require children and adolescents themselves to develop capacities for prevention and self-management. This distinction necessitates interventions that are embedded in students' daily lives and that equip them with sustainable skills to maintain their own psychological well-being.

% % === KEY CONTRIBUTION #1: Research Gap - Elementary Students Understudied ===

A particularly concerning trend is the downward age shift of mental health vulnerabilities in South Korea. In 2023, the proportions of children and adolescents reporting daily life stress reached 56.4\% and 78.9\%, respectively, representing substantial increases from 2020 levels. Most notably, the prevalence of stress among children aged 9--12 increased by 22.1 percentage points, surpassing the 53.1\% rate observed among adolescents in 2020 \citep{moge2023youthsurvey}. This pattern indicates that stress levels once characteristic of middle and high school students have now extended to elementary school populations. Academic pressure arising from intense educational competition has been identified as a primary contributor, with such pressure affecting children as young as four years old through preparation for kindergarten entrance examinations \citep{mknews_2025}.

Despite this expansion of mental health vulnerabilities to younger age groups, elementary school students remain critically understudied in mental health research \citep{weare2011mental, durlak2011impact, salerno2016effectiveness}. Research on childhood mental health has received lower priority relative to adolescent studies, partly because of the practical challenges of collecting survey data from younger children and the historically lower perceived severity of mental health issues in this age group \citep{kiros2025emotional}. As a result, standardized measurement instruments and evidence-based intervention strategies for this population remain limited. This gap is particularly concerning given that stress, poor peer relationships, and low self-esteem---increasingly prevalent among elementary students---are known to diminish life satisfaction and contribute to depression \citep{proctor2009youth}. The present study addresses this gap by focusing specifically on mental health vulnerability among elementary school students.

% % === KEY CONTRIBUTION #2: Incheon as Study Region ==='

Examining how mental health vulnerability manifests across diverse socioeconomic contexts requires the selection of a study region that reflects the heterogeneity of South Korean society. Among the 17 metropolitan cities and provinces, Incheon offers distinct advantages for this purpose. As a metropolitan city within the Seoul metropolitan area, Incheon encompasses considerable regional heterogeneity within a single administrative boundary, including old downtown areas with traditional port facilities, industrial urban zones characterized by manufacturing, newly developed residential areas predominantly inhabited by white-collar professionals, and rural communities comprising multiple islands with limited accessibility.

This diversity provides a microcosm of Korean society, enabling examination of how different socioeconomic and environmental contexts shape students' mental health patterns. Moreover, elementary school students in Incheon exhibit particularly pronounced indicators of mental health vulnerability: 52.8\% report low school life satisfaction, and 71.1\% indicate they are not living an ideal life, rates that substantially exceed national averages \citep{panel2023}. These elevated indicators, combined with the region's socioeconomic diversity, make Incheon a well-suited setting for investigating the relationships among regional context, school environment, and student mental health vulnerability.

% % === KEY CONTRIBUTION #3: School-Level Focus and Hierarchical Modeling ===

Within this regional setting, schools represent the most actionable unit of analysis. Schools constitute the frontline for preventing mental health issues among children and adolescents, providing daily access to students, infrastructure for systematic intervention, and reach to underserved populations \citep{rones2000school}. Schools are especially important for elementary-aged children, as this developmental period represents a critical phase during which social networks and a sense of responsibility are first formally established \citep{eccles2011school}. However, schools operate with limited budgets and resources, necessitating careful identification of school-specific needs for effective intervention. A key insight motivating this study is that schools situated in different environmental contexts are likely to exhibit distinct patterns of student mental health vulnerabilities. These between-school differences reflect the need to identify structural vulnerabilities at the school level as a basis for designing tailored interventions.

This school-level perspective necessitates an analytical approach that can simultaneously model individual student responses and capture systematic school-level patterns. Traditional approaches, including regression models \citep{nordin2009personality} and standard item response theory (IRT) models \citep{Rasch:1961, birnbaum1968some}, present substantial limitations for this purpose. In item response data, both student and item characteristics jointly determine response patterns; regression models using covariates cannot adequately disentangle these dual sources of variation, leading to confounded estimates and potentially biased inferences. IRT addresses some of these limitations by separately modeling latent student abilities and item difficulties within a probabilistic framework. However, IRT-based models structure response probabilities by combining main effects of respondents and items, and thus fail to capture person-item interaction patterns \citep[e.g.,][]{jeon2021mapping, go2022lsirm12pl}. Consequently, these models cannot account for heterogeneity in how specific students respond to specific items \citep{lix2022latent}.

The latent space item response model \citep[LSIRM;][]{jeon2021mapping} addresses these limitations by generating an interaction map that enables interpretation of relationships between respondents and items. This method has been successfully applied in educational settings to analyze students' peer relationships, school activities, depression, and stress \citep[e.g.,][]{kim2022application, park2023social}. However, standard LSIRM models only individual-level item-respondent interactions, presenting a fundamental limitation for our research objectives. When applied to data with hierarchical structure, the method cannot capture the integrated characteristics of groups to which individuals belong. As the number of response items increases and individual tendencies are estimated with greater precision, respondent coordinates in the interaction map increasingly reflect within-group heterogeneity, thereby obscuring between-school differences. Consequently, interpretation becomes limited to relationships among items rather than comparisons of school-level characteristics. Furthermore, post-hoc aggregation of individual latent coordinates to derive school-level representative values yields merely summary statistics of individual estimates, rather than reflecting the mechanism whereby schools are organized around particular orientations and individuals exhibit systematic deviations from these collective tendencies.

% % === PROPOSED SOLUTION AND CONTRIBUTIONS ===

To address these methodological limitations, we propose the hierarchical latent space item response model (HLSIRM), a framework that explicitly integrates hierarchical structure into the model specification. HLSIRM directly infers school-level interaction effects while simultaneously accounting for within-school individual heterogeneity. By specifying item parameters without hierarchical structure, the model maintains measurement invariance, enabling direct comparison of school-level patterns within a unified interaction map. This formulation allows identification of school-specific needs at the item level while preserving individual heterogeneity, thereby supporting the design of targeted interventions.

Applying HLSIRM to elementary school data from Incheon, we aim to numerically and visually identify specific mental health vulnerability domains requiring improvement at each school. To contextualize the model results, we draw on objective indicators of macro-level educational contexts, including regional socioeconomic environments and school infrastructure, to explore potential explanations from multiple angles. This approach leverages publicly available school- and regional-level data, circumventing ethical constraints associated with sensitive individual-level information while enabling contextually informed interpretation of school mental health patterns.

% % === PAPER ORGANIZATION ===

The remainder of this paper is organized as follows. Section \ref{sec:data} presents detailed information about the collected data and response characteristics. Section \ref{sec:model} provides the structure of HLSIRM and implementation details. Section \ref{sec:application} presents the results of applying the proposed model to actual data. Section \ref{sec:conclusion} concludes the paper with discussion of implications and future directions.

\section{Data: Mental Health Vulnerability Survey of Elementary Schools in Incheon}\label{sec:data}

This section describes the data used to examine mental health vulnerability among elementary school students in Incheon, South Korea. We first describe the sampling design and regional classification, then present the survey instrument and measurement characteristics, and finally provide preliminary evidence of school-level heterogeneity that motivates the hierarchical modeling approach developed in Section~\ref{sec:model}.

\subsection{Study Region and Sampling Design}\label{sec:region}

To operationalize Incheon's socioeconomic diversity for systematic analysis, we classified its ten administrative districts into four regional types. Table~\ref{tab:table1} presents the resulting classification along with the sample size, change in Gross Regional Domestic Product (GRDP), and elementary school enrollment for each district.

\begin{table}[hbtp]
\centering
\small
\renewcommand{\arraystretch}{1.15}
\begin{tabular}{llrrr}
Regional Type & District Name & Sample Size & GRDP & Elementary school enrollment \\
\hline
\multirow{2}{*}{Old Downtown Areas (D)}
  & Dong-gu     & 186  & -0.1 & 3,109 \\
  & Michuhol-gu & 226 & -2.0 & 15,980 \\
\hline
\multirow{3}{*}{Industrial Urban Areas (I)}
  & Bupyeong-gu & 194 & -0.8 & 21,225 \\
  & Gyeyang-gu  & 208 & -1.0 & 10,758 \\
  & Namdong-gu  & 225 &  3.1 & 24,192 \\
\hline
\multirow{3}{*}{Newtown Areas (N)}
  & Seo-gu      & 304 &  4.6 & 37,878 \\
  & Jung-gu     & 335 &  2.5 & 9,669  \\
  & Yeonsu-gu   & 306 &  6.4 & 25,598 \\
\hline
\multirow{2}{*}{Rural Island Areas (R)}
  & Ganghwa-gun & 174 & -5.6 & 1,936 \\
  & Ongjin-gun  & 52  &  0.6 & 387   \\
\end{tabular}

\caption{
Regional classification of Incheon administrative districts with student sample sizes, change in Gross Regional Domestic Product (GRDP) between 2011 and 2019 (\%), and elementary school enrollment as of 2024.
}
\label{tab:table1}
\end{table}

Each regional type represents distinct socioeconomic conditions that may differentially influence student mental health patterns. Old downtown areas~(D) are characterized by aging infrastructure, traditional port-based commerce, and relatively older resident populations, with established but resource-limited school systems. Industrial urban areas~(I) feature manufacturing and commercial activities, working-class family compositions, and moderate population density; schools in these areas may face challenges related to parental work schedules and economic pressures on families. Newtown areas~(N) represent recently developed residential zones predominantly inhabited by white-collar professionals, characterized by modern infrastructure, higher socioeconomic status, and intense academic competition. Rural island areas~(R) comprise geographically isolated communities with limited transportation accessibility, sparse populations, and restricted access to educational and welfare resources.

This four-type classification guided the sampling design for student data collection. Data were obtained from the Longitudinal Study Supporting the Development of a Student Growth and Adaptation System (2022--2023), which surveyed 2,210 elementary school students across 35 schools in Incheon \citep{kim2022_data, kim2023_data}. Sample sizes across schools were designed to reflect actual proportions of students by regional type, resulting in an unbalanced design due to the use of cluster (classroom) census sampling. This imbalance is particularly pronounced in rural areas, which consist exclusively of island communities with limited transportation accessibility; student sample sizes differ by up to 3.8-fold between schools, with rural areas having substantially smaller samples. The names and detailed sample sizes of the 35 selected schools are provided in Section~1.1 of the Supplementary Material. Regional-level contextual information, including economic characteristics, educational infrastructure, and environmental differences, is summarized in Sections~2.1 and~2.2 of the Supplementary Material.

\subsection{Survey Instrument and Item Characteristics}\label{sec:instrument}

The measurement instrument comprises 82 items assessing elementary students' mental health vulnerability, focusing specifically on students' self-assessed psychological states and behavioral patterns. Items not directly related to student mental health, including teacher attitude assessments and parenting approach indicators, were excluded to maintain conceptual coherence and to focus the analysis on student-level factors amenable to school-based intervention.

The 82 items are organized into 12 categorical domains that capture distinct dimensions of mental health vulnerability among elementary students. Table~\ref{Tab:Table2} presents these domains with their corresponding item indices.

\begin{table}[hbtp]
\centering
\footnotesize
\renewcommand{\arraystretch}{1.15}

\begin{tabular}{cll}
Category & Description & Item Index \\ \hline
SH  & Poor study habits reflecting low self-regulation and academic interest 
    & 1--10  \\
DH & Unhealthy daily habits associated with reduced wellbeing
    & 11--22 \\
CE  & No counseling experience for seeking help and utilizing mental health support
    & 23--24 \\
SIV & Low involvement in school activities indicating weaker social connectedness
    & 25--27 \\
SAD & Difficulties in school adjustment reflecting impaired psychosocial function
    & 28--38 \\
HP  & Low happiness and life satisfaction
    & 39--41 \\
FD  & Anxiety and depression symptom
    & 42--47 \\
ST  & Stress (grade, coursework, career, relationship, etc.)
    & 48--55 \\
EM  & Difficulties in emotion regulation, stress coping, and positive attitude maintenance
    & 56--64 \\
SS  & Low social competence for peer cooperation and understanding
    & 65--76 \\
SPD & Smartphone dependency associated with anxiety and loneliness
    & 77--80 \\
DGT & Digital device usage for academic and non-academic purpose
    & 81--82 \\
\end{tabular}

\caption{
Twelve categorical domains of mental health vulnerability with brief descriptions and corresponding item indices. Each domain captures a distinct dimension of student psychological functioning; detailed item descriptions are provided in Section~1.2 of the Supplementary Material.
}
\label{Tab:Table2}
\end{table}

These 12 domains collectively operationalize the concept of mental health vulnerability introduced in Section~\ref{sec:introduction}. SH and DH assess foundational self-regulation competencies essential for maintaining psychological stability. CE captures help-seeking behavior and utilization of mental health support, while SIV and SAD measure disengagement from the school environment. HP, FD, and ST directly assess current psychological states. EM and SS capture vulnerabilities in emotional regulation and social coping under psychological distress or collaborative classroom demands. Finally, SPD and DGT address contemporary digital factors increasingly relevant to youth mental health.

The survey employed a standardized 5-point Likert scale for severity-based assessments and a binary response format for experience-based items. All responses were binarized according to a predetermined coding scheme designed to identify mental health vulnerability. Responses at scale points 4 and 5, indicating problematic functioning, were coded as 1 to denote vulnerability, while responses at scale points 1 through 3 were coded as 0 to denote the absence of vulnerability. The DGT category required separate treatment due to its 7-point measurement structure: responses at scale points 5 through 7, representing extended engagement with digital devices, were coded as 1 to indicate potentially problematic usage patterns.

Comprehensive descriptions of survey questions for each item are provided in Section~1.2 of the Supplementary Material. Descriptive statistics of item responses at the school and regional levels are reported in Section~1.3 of the Supplementary Material, where notable between-school and between-region variation is observed for several items (e.g., SH05--SH07, CE01--CE02, and SIV01--SIV03).

%\textcolor{red}{[provide descriptive statistics at the school level and regional level}
%\textcolor{orange}{add in Supplementary 1.3}

Beyond the student-level survey, we obtained school-level institutional data characterizing each school's extracurricular educational environment. These covariates include the annual hours devoted to school bullying prevention and daily life safety education programs, both of which are systematically documented by educational authorities and publicly available \citep{schoolsafe2025, schoolbul2025}. School bullying prevention programs typically address peer relationship skills, conflict resolution, and supportive school climates, while daily life safety education covers hygiene, health awareness, and healthy lifestyle habits. These institutional variables serve as contextual information for validating whether school positions in the latent space align with known educational interventions (Section~\ref{sec:interaction_map}). The specific values and detailed descriptions of these covariates are presented in Section~2.3 of the Supplementary Material.

\subsection{Preliminary Evidence for School-Level Heterogeneity} \label{sec:preliminary}

Before introducing the hierarchical modeling framework in Section \ref{sec:model}, we present preliminary evidence of systematic school-level variation in mental health vulnerability patterns. This variation motivates the need for an analytical approach that can capture both individual-level responses and school-level structures simultaneously.

\paragraph*{School-Level Response Patterns}

\begin{figure}[htb]
    \centering
    \includegraphics[width=0.9\textwidth,height=9cm]{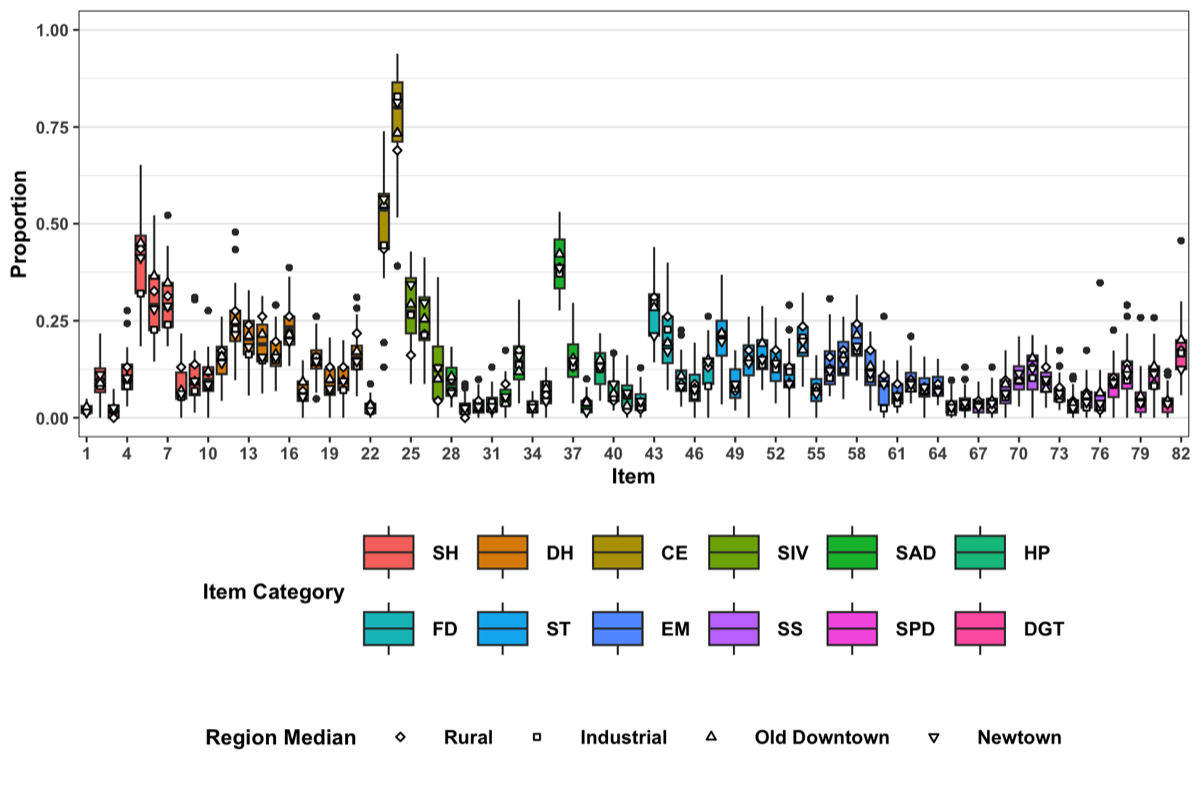}
    \caption{
    Distribution of school-level vulnerability response proportions across all 82 items. Each box represents the distribution of proportions across 35 schools for a single item; wider boxes indicate greater between-school heterogeneity. Regional median values are indicated by symbols (see legend).
    }
    \label{fig:boxplot}
\end{figure}

Figure \ref{fig:boxplot} presents the distribution of vulnerability response proportions across schools for all 82 items. The substantial variation in box widths and positions indicates that schools differ meaningfully in their students' mental health profiles. Items 23, 24, and 36---related to the absence of counseling experience and low preference for consulting teachers---exhibit notably high vulnerability response rates across schools.

\begin{figure}[htb]
    \centering
    \subfloat[Counseling Experience (CE)]{{\includegraphics[width=0.5\textwidth,height=5.5cm]{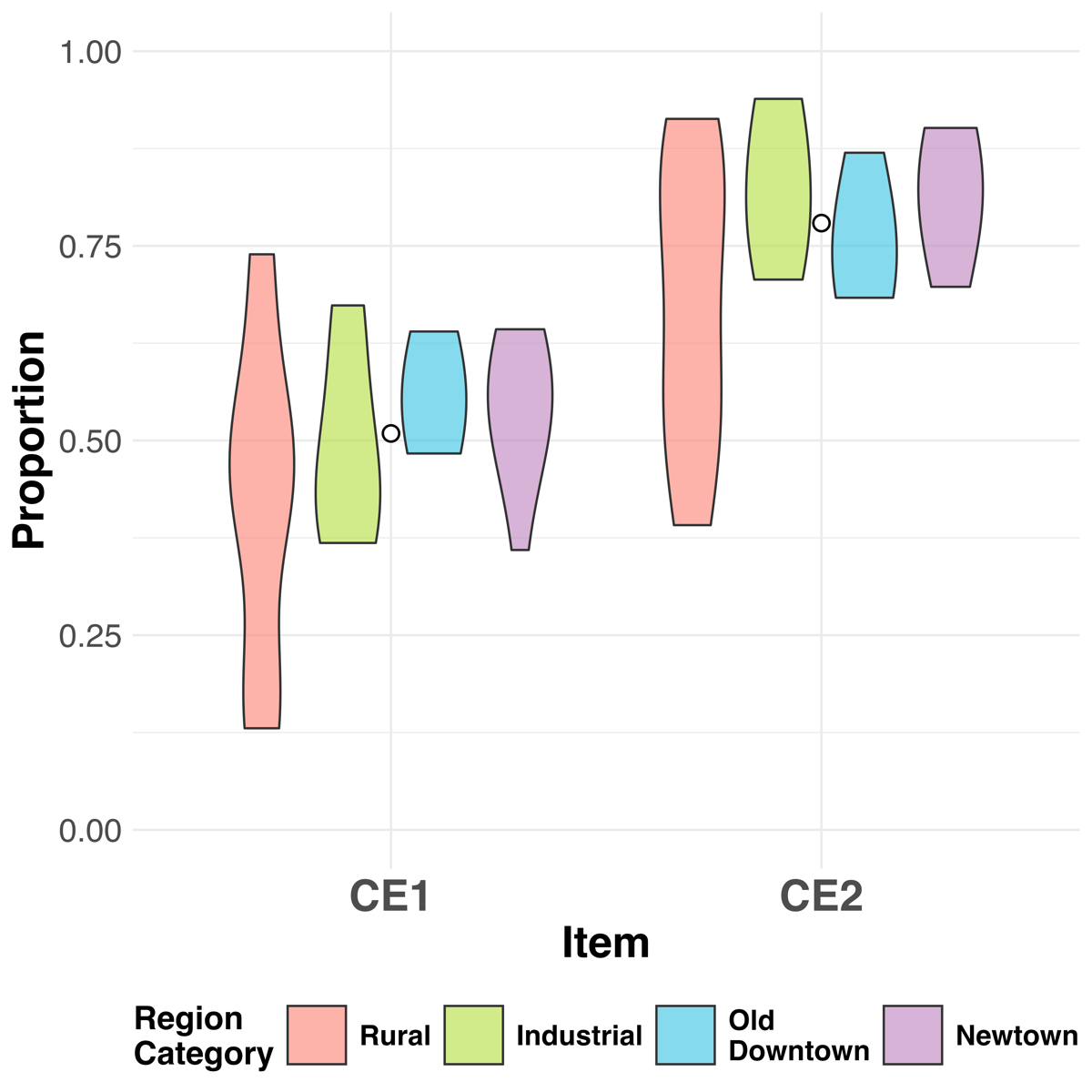}} \label{fig:CE}}
    \subfloat[School Adaptation (SAD)]{{\includegraphics[width=0.5\textwidth,height=6cm]{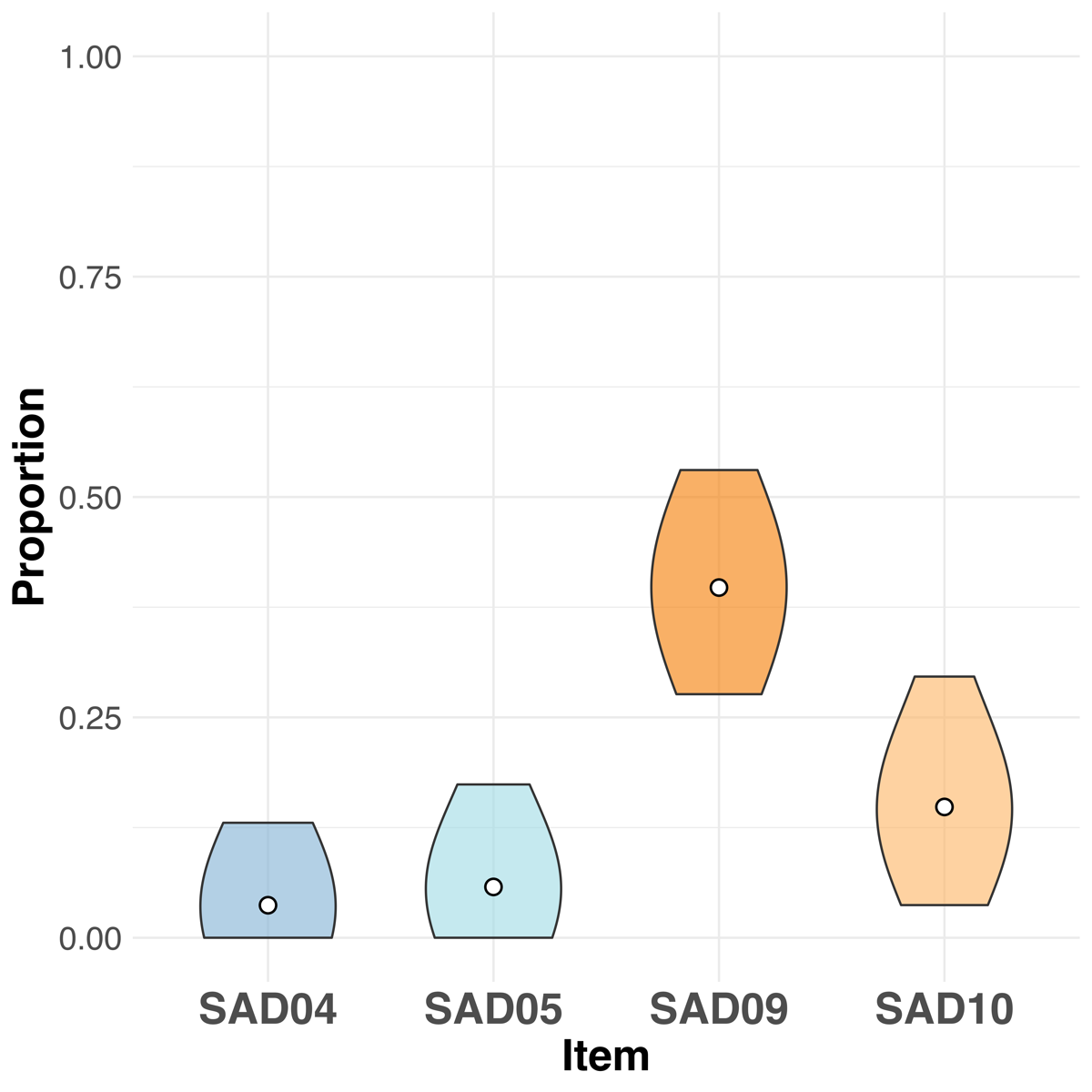}} \label{fig:SAD}}
    \caption{
    School-level vulnerability response rates by (a) regional type for the Counseling Experience (CE) category and (b) the School Adaptation (SAD) category. Each data point represents an individual school's mean vulnerability response rate for a given item, colored by regional type. Asterisks indicate the overall mean for each item.
    }
    \label{fig:scatter}
\end{figure}

Figure~\ref{fig:scatter} provides a more detailed examination of school-level variation for selected item categories. Each data point represents the mean proportion of vulnerability responses for an individual school, differentiated by regional type: blue for old downtown areas~(D), green for industrial urban areas~(I), purple for newtown areas~(N), and red for rural island areas~(R). Asterisks in the violin plots indicate the overall mean vulnerability response rate for each item.

For the Counseling Experience category shown in Figure~\ref{fig:CE}, the absence of counseling experience with professional counselors (CE02) consistently shows higher rates than the absence of homeroom teacher counseling (CE01) across all regional types. Rural areas exhibit greater variability than other regions, likely reflecting heterogeneity in counseling resource availability across island communities. This pattern aligns with the School Adaptation findings in Figure~\ref{fig:SAD}, where a high proportion of students express unwillingness to consult teachers when experiencing personal concerns (SAD09), contrasting with the low proportions observed for items related to peer relationship difficulties (SAD04, SAD05). Notably, the proportion of students reporting discomfort in talking with teachers (SAD10) is relatively low, suggesting that the quality of teacher--student relationships may be distinct from students' willingness to seek counseling---a nuanced distinction that item-level analysis can reveal.

In summary, the preliminary analyses presented above reveal meaningful between-school heterogeneity in mental health vulnerability at the item level. Three features of this data structure are particularly salient. First, between-school variation in vulnerability patterns is systematic rather than random, but this variation is obscured when only individual-level parameters are estimated. Second, substantial local dependence among items necessitates modeling approaches that capture item-respondent interaction patterns beyond main effects. Third, the nested structure of students within schools represents a substantively important feature that should be explicitly incorporated into the analytical framework. These observations motivate the hierarchical latent space item response model developed in the following section, which is designed to simultaneously accommodate school-level variation and item-level interaction patterns within a unified framework.

\section{Hierarchical Latent Space Item Response Model}\label{sec:model}

This section introduces the hierarchical latent space item response model (HLSIRM), a framework that enables investigation of item-level heterogeneity at both the student and school levels. We first specify the model and justify the inner product formulation, then describe the hierarchical structure for respondent parameters, and finally present the prior distributions and MCMC estimation procedure.

\subsection{Hierarchical Latent Space Item Response Model}\label{sec:hlsirm}

\subsubsection{Model Specification}

Let $y_{ij(k)}$ denote the binary response of respondent $i \in \{1, \ldots, n_k\}$ nested within school $k \in \{1, \ldots, K\}$ to item $j \in \{1, \ldots, p\}$. In our application, respondents are elementary school students, schools represent the higher-level grouping structure, and items assess various dimensions of mental health vulnerability. The school-specific response matrix $\mathbf{Y}_{(k)} \in \{0, 1\}^{n_k \times p}$ contains all responses for school $k$, where $y_{ij(k)} = 1$ indicates a vulnerability response and $y_{ij(k)} = 0$ indicates adequate functioning.

The probability of observing a vulnerability response is modeled as:
\begin{equation}\label{eq:main}
\text{logit} \left( \Pr(y_{ij(k)} = 1 \mid \alpha_{i(k)}, \beta_j, \mathbf{z}_{i(k)}, \mathbf{w}_j) \right) = \alpha_{i(k)} + \beta_j + \mathbf{z}_{i(k)}^\top \mathbf{w}_j + \varepsilon_{ij(k)},
\end{equation}
where each component captures a distinct aspect of the response process.

\begin{itemize}
\item {\bf Main Effect Parameters}: The individual-specific intercept $\alpha_{i(k)} \in \mathbb{R}$ represents the baseline vulnerability of student $i$ within school $k$, capturing systematic individual differences in the propensity to report mental health vulnerabilities independently of specific item content or interaction effects. Students with higher values of $\alpha_{i(k)}$ exhibit greater overall vulnerability across all items. The item-specific intercept $\beta_j \in \mathbb{R}$ quantifies the baseline endorsement rate of item $j$ across all respondents. Items with higher values of $\beta_j$ are more frequently endorsed as vulnerabilities, indicating either greater prevalence in the population or a lower item threshold. Critically, $\beta_j$ remains constant across all schools to ensure measurement invariance, allowing meaningful comparisons of school-level patterns.

\item {\bf Interaction Effect - The Inner Product Term}: The inner product $\mathbf{z}_{i(k)}^\top \mathbf{w}_j$ captures the interaction between student $i$ in school $k$ and item $j$ through their respective latent position vectors in a $D$-dimensional space. This term represents the core innovation inherited from the LSIRM framework: the ability to model respondent-item interactions beyond additive main effects. The student position vector $\mathbf{z}_{i(k)} \in \mathbb{R}^D$ represents the latent profile of student $i$ within the interaction map, while the item position vector $\mathbf{w}_j \in \mathbb{R}^D$ captures the dimensions along which the item discriminates among students.

The inner product formulation has a natural geometric interpretation. The angle between $\mathbf{z}_{i(k)}$ and $\mathbf{w}_j$ determines the type of interaction: small angles indicate positive interaction (elevated probability of endorsing the item as a
vulnerability), perpendicular orientations indicate independence, and opposing directions indicate negative interaction (reduced probability of endorsement). Vector magnitude reflects interaction strength: larger student vector magnitudes indicate more strongly differentiated response patterns, while larger item vector magnitudes indicate stronger discrimination among students. Items with small magnitudes contribute little to the interaction structure and function primarily through their main effects.

\item {\bf The Error Term}: The error term $\varepsilon_{ij(k)} \overset{\text{i.i.d.}}{\sim} \text{N}(0, 1)$ prevents the interaction map from overfitting idiosyncratic response patterns. Without this term, the model would attempt to explain all response variation through the systematic components, potentially distorting the interaction map to accommodate noise. The inclusion of $\varepsilon_{ij(k)}$ allows for case-specific residual variability, ensuring that the estimated interaction structure reflects genuine patterns rather than sampling artifacts. This specification follows \citet{hoff2021additive} for additive and multiplicative effects models.
\end{itemize}

\subsubsection{Rationale for the Inner Product Formulation}

The original LSIRM employs Euclidean distance ($-\gamma\|\mathbf{z}_{i(k)} - \mathbf{w}_j\|$) to capture respondent-item interactions, where proximity indicates strong association \citep{go2022lsirm12pl}. In HLSIRM, we adopt the inner product formulation for three reasons \citep{hoff2007modeling}.

First, the Euclidean distance formulation imposes a monotonic relationship between proximity and interaction strength, restricting all associations to positive homophily. This constraint is inappropriate for mental health assessment data, where both positive and negative associations between student profiles and items are expected. The inner product naturally accommodates both through vector alignment \citep{hoff2005bilinear}.

Second, the distance-based formulation conflates pattern similarity with response intensity. Two students equidistant from an item have identical predicted interactions regardless of whether they occupy high-magnitude or low-magnitude positions. The inner product decouples direction from magnitude, allowing the model to distinguish between students with strong versus weak associations even when their directional profiles are similar. This distinction is substantively important for identifying students with pronounced vulnerability patterns versus those with more moderate tendencies.

Third, the inner product formulation offers computational advantages. Euclidean distances remain invariant under translation, rotation, and reflection, introducing identifiability challenges that require Procrustes matching and may hinder MCMC convergence. The inner product formulation requires addressing only rotational invariance, leading to more stable posterior estimation and reduced computational complexity \citep{minhas2019inferential}.

\subsection{Hierarchical Structure for Respondent Parameters}\label{sec:hierarchical}

\begin{figure}[hbt!]
\centering
\includegraphics[width=1.0\textwidth, height=10cm]{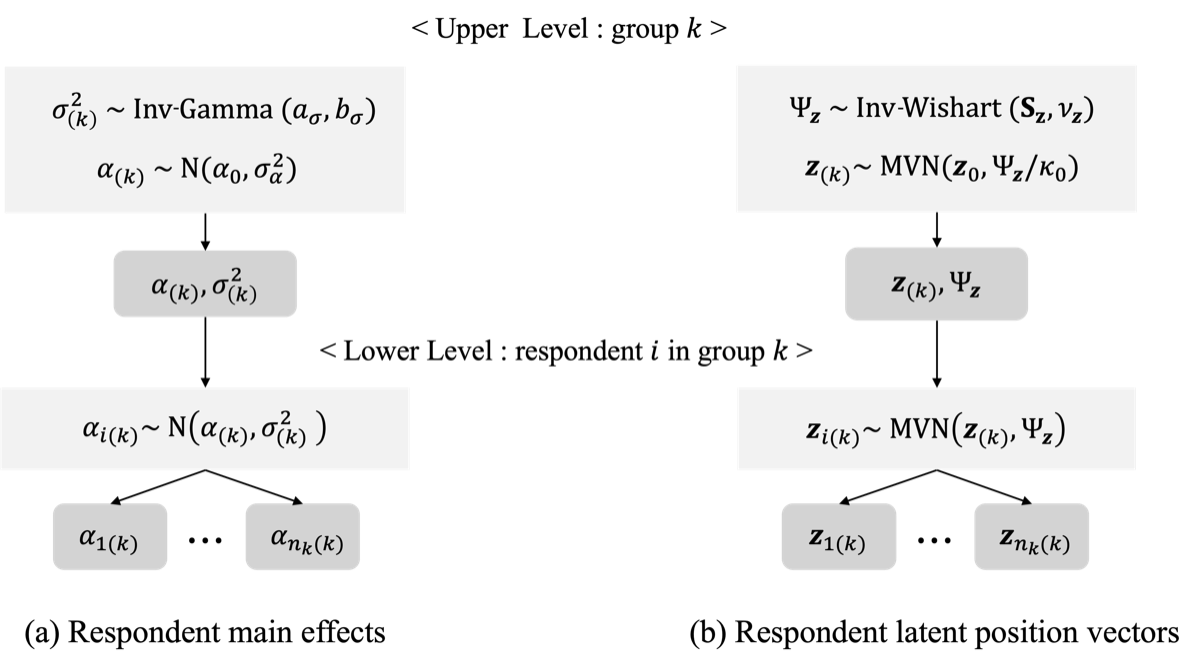}
\caption{
Hierarchical structure of respondent parameters in HLSIRM. Panel~(a) illustrates the nested structure where individual student intercepts $\alpha_{i(k)}$ vary around school-level means $\alpha_{(k)}$. Panel~(b) illustrates the analogous structure for latent position vectors, where individual positions $\mathbf{z}_{i(k)}$ vary around school-level positions $\mathbf{z}_{(k)}$.
}
\label{fig:hier_structure}
\end{figure}

The distinguishing feature of HLSIRM is its explicit hierarchical specification for respondent parameters, which enables identification of school-level patterns while preserving individual-level heterogeneity. This structure is depicted schematically in Figure \ref{fig:hier_structure}.

\subsubsection{Upper-Level Parameters: School Characteristics} 

At the upper level, each school $k$ is characterized by two parameters representing the central tendency of its students. The school-level intercept $\alpha_{(k)} \in \mathbb{R}$ represents the average baseline vulnerability of school $k$, summarizing the collective tendency of students within the school to report mental health vulnerabilities. Schools with higher values of $\alpha_{(k)}$ have student bodies that exhibit greater overall vulnerability across all items.

The school-level latent position vector $\mathbf{z}_{(k)} \in \mathbb{R}^D$ represents the aggregate interaction profile of school $k$ within the interaction map. This vector summarizes the average tendency of students within the school to interact with items in
particular ways: schools positioned close to certain item vectors have student bodies that systematically exhibit vulnerabilities in those domains, while schools positioned opposite to item vectors demonstrate adequate functioning in those domains.

\subsubsection{Lower-Level Parameters: Individual Student Characteristics}

At the lower level, individual students are characterized by parameters that deviate from their respective school means. The individual intercept $\alpha_{i(k)}$ is drawn from a school-specific distribution centered on the school-level intercept:
\begin{equation}
\alpha_{i(k)} \mid \alpha_{(k)}, \sigma_{(k)}^{2} \sim \text{N}(\alpha_{(k)}, \sigma_{(k)}^{2}), \quad \text{for } i = 1, \ldots, n_k.
\end{equation}
The variance parameter $\sigma_{(k)}^2$ captures within-school heterogeneity in baseline vulnerability. Schools with larger $\sigma_{(k)}^2$ exhibit greater diversity among their students' overall mental health patterns, while schools with smaller $\sigma_{(k)}^2$ have more homogeneous student bodies.

Similarly, the individual latent position vector $\mathbf{z}_{i(k)}$ is drawn from a school-specific multivariate distribution centered on the school-level position \citep{hoff2011hierarchical}:
\begin{equation}
\mathbf{z}_{i(k)} \mid \mathbf{z}_{(k)}, \boldsymbol{\Psi}_{\mathbf{z}} \sim \text{MVN}(\mathbf{z}_{(k)}, \boldsymbol{\Psi}_{\mathbf{z}}), \quad \text{for } i = 1, \ldots, n_k.
\end{equation}
The covariance matrix $\boldsymbol{\Psi}_{\mathbf{z}}$ captures within-school heterogeneity in interaction profiles. Schools with larger dispersion in $\boldsymbol{\Psi}_{\mathbf{z}}$ have students whose interaction patterns vary substantially, while schools with smaller dispersion have students who respond to items in more consistent ways.

% \textcolor{red}{[do we estimate $\kappa$? why ther eare many sigma, instead of one, at the population level? ]}

% \textcolor{blue}{[$\kappa$를 추정할까요?: 아니요, 1로 고정합니다. 이는 응용프로그램에 명시되어 있습니다. 이 섹션에서는 카파가 일반적으로 어떻게 작동하는지 간략하게 설명했습니다.]}

% \textcolor{blue}{[모집단 수준에서 시그마가 하나가 아닌 여러 개인 이유는 무엇일까요?: 시그마가 $k$개인 이유는 계층 구조 때문입니다. 잠재 공간(분산 및 경향)에서 응답자의 분포가 학교마다 다른 것처럼, 응답자의 주효과(능력)도 학교마다 다르다고 가정합니다. 따라서 각 학교마다 $\alpha_{i(k)}$를 표본 추출하기 위한 분산을 다르게 조정했습니다.]}

\subsection{Prior Distributions}\label{sec:priors}

We employ a fully Bayesian framework with conjugate and weakly informative prior distributions that enable stable posterior inference while allowing the data to drive parameter estimates.

\paragraph*{Upper-Level Priors for Respondent Parameters} The school-level intercepts are drawn from a common population distribution:
\begin{equation}
\alpha_{(k)} \mid \sigma_{\alpha}^{2} \sim \text{N}(\alpha_0, \sigma_{\alpha}^{2}), \quad \text{for } k = 1, \ldots, K,
\end{equation}
where $\alpha_0$ represents the global mean intercept across all schools, allowing schools to deviate from the population mean while maintaining regularization toward the global average. The school-level latent position vectors follow:
\begin{equation}
\mathbf{z}_{(k)} \mid \boldsymbol{\Psi}_{\bf z}\sim \text{MVN}(\mathbf{z}_0, \boldsymbol{\Psi}_{\bf z}/\kappa_0), \quad \text{for } k = 1, \ldots, K,
\end{equation}
where $\mathbf{z}_0$ represents the global mean position in the interaction map. This hierarchical prior on the latent positions follows the modeling strategy of hierarchical multilinear models \citep{hoff2011hierarchical}. We set $\alpha_0 = 0$ and $\mathbf{z}_0 = \mathbf{0}$ to center the interaction map at the origin and $\kappa_0 = 1$ for a diffuse prior on the school-level positions.

\paragraph*{Lower-Level Priors for Respondent Variance Parameters} The within-school variance parameters receive inverse-gamma priors:
\begin{equation}
\sigma_{(k)}^{2} \sim \text{Inv-Gamma}(a_\sigma, b_\sigma),
\end{equation}
and the within-school covariance matrices receive inverse-Wishart priors:
\begin{equation}
\boldsymbol{\Psi}_{\mathbf{z}} \sim \text{Inv-Wishart}(\mathbf{S}_\mathbf{z}, \nu_\mathbf{z}).
\end{equation}
These conjugate specifications facilitate efficient Gibbs sampling  for variance and covariance parameters while accommodating school-specific heterogeneity patterns.

\paragraph*{Item Parameter Priors} Item parameters are specified without hierarchical structure to preserve measurement invariance across schools \citep{luo2023bayesian, kang2024recent}. The item intercepts follow:
\begin{equation}
\beta_j \sim \text{N}(\beta_0, \tau^2), \quad \text{for } j = 1, \ldots, p,
\end{equation}
where $\beta_0 = 0$ represents the prior mean and $\tau^2$ is fixed to ensure identifiability.  The item latent position vectors follow:
\begin{equation}
\mathbf{w}_j \mid \boldsymbol{\Psi}_{\bf w} \sim \text{MVN}(\mathbf{w}_0, \boldsymbol{\Psi}_{\bf w}), \quad \text{for } j = 1, \ldots, p,
\end{equation}
where $\mathbf{w}_0 = \mathbf{0}$ centers items at the origin and $\boldsymbol{\Psi}_{\bf w} \sim \text{Inv-Wishart}(\mathbf{S}_{\bf w}, \nu_{\bf w})$ allows flexible modeling of item latent position dispersion.

\subsection{MCMC Sampling}\label{sec:mcmc}

Parameter estimation proceeds via Markov chain Monte Carlo (MCMC) methods within the Bayesian framework. The vulnerability response probability for school $k$ can be expressed in matrix form as:
\begin{equation}
\text{logit}(\boldsymbol{\Theta}_{(k)}) = \boldsymbol{\alpha}_{(k)} \mathbf{1}_p^\top + \mathbf{1}_{n_k} \boldsymbol{\beta}^\top + \mathbf{Z}_{(k)}\mathbf{W}^\top + \mathbf{E}_{(k)},
\end{equation}
where $\boldsymbol{\Theta}_{(k)}$ is the $n_k \times p$ matrix of response probabilities, $\mathbf{Z}_{(k)}$ is the $n_k \times D$ matrix of student position vectors, $\mathbf{W}$ is the $p \times D$ matrix of item position vectors, and $\mathbf{E}_{(k)}$ is the matrix of error terms. The bilinear term $\mathbf{Z}_{(k)}\mathbf{W}^\top$ provides a low-rank factorization of the school-specific response surface, structurally analogous to the eigenmodel framework \citep{hoff2007modeling, salak2008bayes}.

HLSIRM updates all parameters ($\alpha_{(k)}, \alpha_{i(k)}, \beta_j, \mathbf{z}_{(k)}, \mathbf{z}_{i(k)}, \mathbf{w}_j$) from their posterior distributions during each iteration. Rather than making separate acceptance decisions for individual parameters, the Metropolis-Hastings algorithm \citep{metropolis1953equation, hastings1970monte} determines acceptance once per iteration for each school $k$ based on the jointly proposed update of the combined parameter set. This joint updating procedure enhances computational efficiency while improving posterior exploration, and appropriately captures the interdependencies between $\alpha_{(k)}$ and $\alpha_{i(k)}$, as well as between $\mathbf{z}_{(k)}$ and $\mathbf{z}_{i(k)}$.

\paragraph*{Addressing Rotational Invariance} The inner product formulation introduces rotational invariance: if $\mathbf{R}$ is any orthogonal matrix, then $\mathbf{z}_{i(k)}^\top \mathbf{w}_j = (\mathbf{R}\mathbf{z}_{i(k)})^\top (\mathbf{R}\mathbf{w}_j)$. This invariance means that the likelihood is unchanged under simultaneous rotation of all position vectors, creating a non-identifiability that must be resolved for meaningful interpretation.

We address this through post-processing Procrustes alignment \citep{gower1975generalized, friel2016interlocking}. After MCMC sampling, posterior samples are aligned to a reference configuration by solving the orthogonal Procrustes problem that minimizes the squared Frobenius norm between the reference and transformed configurations. This alignment is applied to all position vectors, including both student and item vectors, enabling meaningful statistical inference and consistent visualization across posterior samples. The complete MCMC sampling algorithm, including conditional posterior distributions, convergence diagnostics, and the Procrustes matching procedure, is provided in Sections~3 and~4 of the Supplementary Material.

\paragraph*{Interaction Map Dimensionality} While HLSIRM permits arbitrary specification of the interaction map dimension $D$, we fix $D = 2$ throughout our application for two reasons \citep{hoff2002latent, dangelo2019latent}. First, two-dimensional representations enable direct visualization of school-item relationships, which is essential for communicating findings to practitioners and policymakers. Second, higher-dimensional spaces increase computational complexity and interpretation difficulty without necessarily improving model fit for the patterns of interest. The primary purpose of the interaction map is not to optimize prediction but to enable substantive interpretation of relationship patterns that would otherwise remain inaccessible.

\section{Real Data Application}\label{sec:application}

This section presents the results of applying HLSIRM to the Incheon elementary school mental health vulnerability data described in Section~\ref{sec:data}. We first describe the analysis implementation and evaluate model fit, then present results organized around three levels of interpretation: main effect parameters that characterize overall vulnerability (Section~\ref{sec:main_effects}), the interaction map that reveals relationships among items and schools (Section~\ref{sec:interaction_map}), and school-specific patterns that inform targeted intervention design (Section~\ref{sec:intervention}).

\subsection{Analysis Implementation}\label{sec:analysis}

HLSIRM was applied to the complete dataset of 2,210 students across 35 schools described in Section~\ref{sec:data}. MCMC sampling was conducted for 30,000 iterations, with the first 5,000 discarded as burn-in. The remaining samples were thinned by retaining every fifth iteration, yielding 5,000 posterior samples for inference. The interaction map dimension was fixed at $D = 2$ as discussed in Section~\ref{sec:mcmc}.

Hyperparameters were specified to be weakly informative while ensuring stable estimation. We set $\alpha_0 = \beta_0 = 0$ and $\mathbf{z}_0 = \mathbf{w}_0 = \mathbf{0}$ to center the interaction map at the origin. The variance parameters for $\alpha_{(k)}$ and $\beta_j$ were fixed at $\sigma_{\alpha} = \tau = 2.5$. For the inverse-Wishart distributions governing position vector covariances, degrees of freedom were set to $\nu_\mathbf{z} = \nu_\mathbf{w} = D + 1$ to maintain minimal informativeness, with scale matrices $\mathbf{S}_\mathbf{z} = \mathbf{S}_\mathbf{w} = 2 \cdot \mathbf{I}$ \citep{schuurman2016comparison}. We set $\kappa_0 = 1$ for a diffuse prior \citep{hoff2011hierarchical}. The inverse-gamma priors for within-school variance parameters
were specified with $a_\sigma = b_\sigma = 1$, ensuring that $\sigma_{(k)}^2$ spans a broad range without concentrating near zero \citep{gelman2006prior}. The error variance $1/\phi$ was fixed at 1 because $\phi$ is confounded with the magnitude of other parameters \citep{hoff2007model}. Sensitivity analyses examining alternative prior specifications are reported in Section~5 of the Supplementary Material and confirm the robustness of our findings.

\subsection{Model Fit Evaluation}\label{sec:modelfit}

Before examining substantive results, we assess model adequacy through posterior predictive checking and classification performance evaluation.

%\textcolor{red}{[consider move this to the beginning area (of the application section) or to the supplement]}
% Section 4.5에서 4.2로 이동하였습니다. Section 4 도입에서 구현 설명 -> 적합 평가 -> 실제 데이터 적용 해석 순으로 진행함을 언급하였습니다. Supplementary Material로 이동할 경우 학교별 performance table이 속한 Section 10 앞에 위치할 예정입니다. 

\subsubsection{Posterior Predictive Checking}

We generated 200 replicated datasets from the estimated parameters to assess whether HLSIRM adequately captures the data-generating process. Individual intercepts were sampled from school-specific distributions centered on $\hat{\alpha}_{(k)}$ with variance $\hat{\sigma}_{(k)}^2$, and position vectors were sampled from distributions centered on $\hat{\mathbf{z}}_{(k)}$ with covariance $\hat{\boldsymbol{\Psi}}_{\mathbf{z}}$.

\begin{figure}[htb]
    \centering
    \subfloat[Simulated $\alpha_{i(k)}^{[s]}$]{{\includegraphics[width=0.4\textwidth]{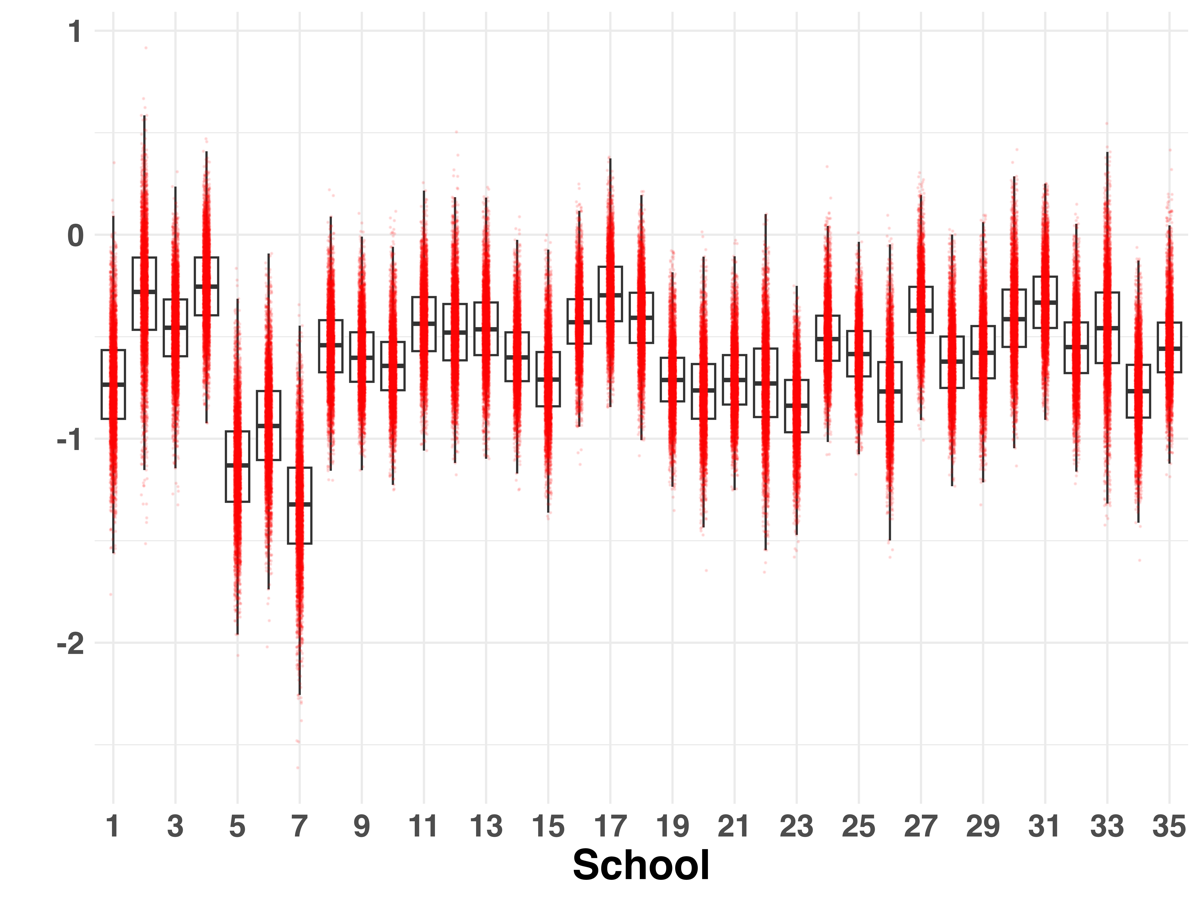}}\label{fig:sim1 alpha}}
    \subfloat[Simulated $\beta_{j(k)}^{[s]}$]{{\includegraphics[width=0.4\textwidth]{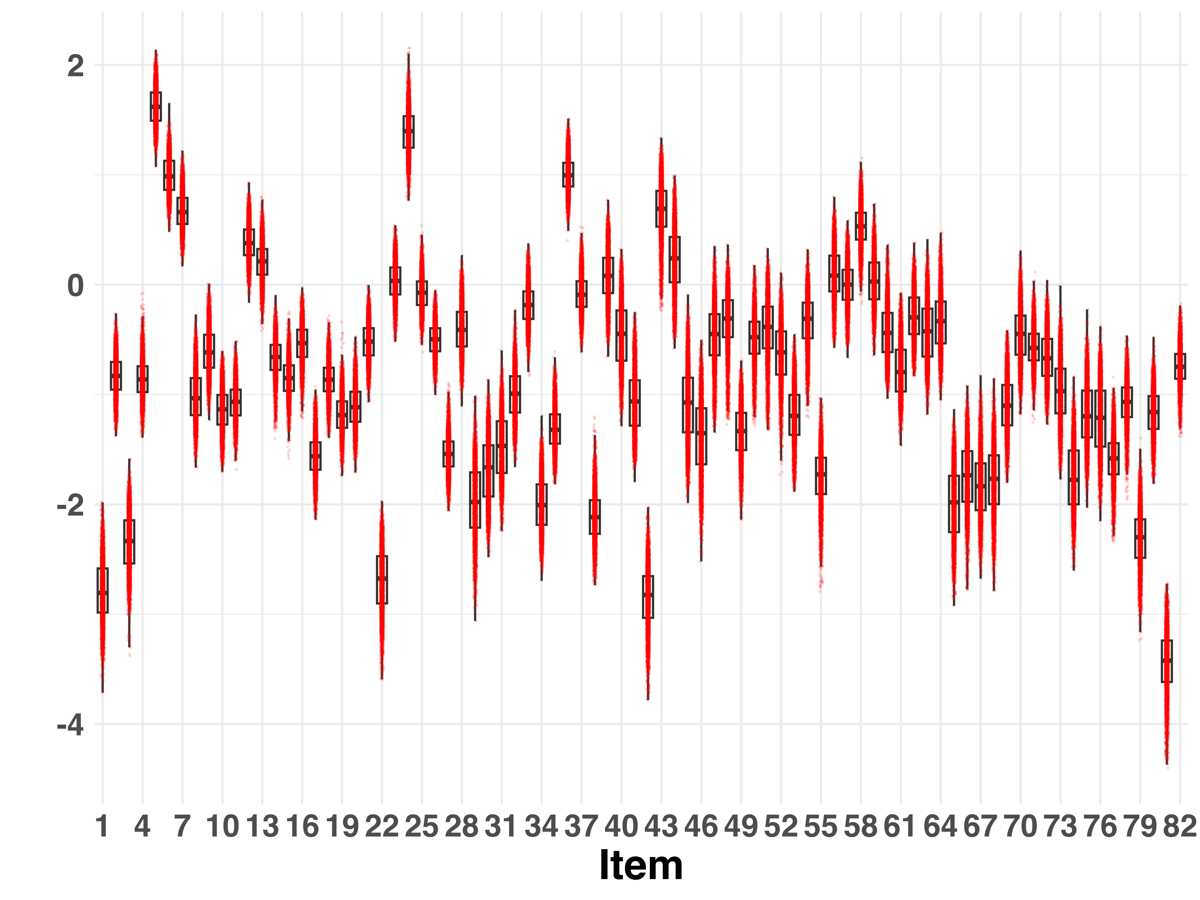}}\label{fig:sim1 beta}}
    \caption{
    Parameter recovery assessment via posterior predictive simulation ($S = 200$ replications). Boxplots show distributions of parameters simulated from the estimated hierarchical structure; red dots indicate the original posterior mean estimates. Panel~(a) displays individual-level intercepts $\alpha_{i(k)}^{[s]}$ simulated from $\text{N}(\hat{\alpha}_{(k)}, \hat{\sigma}_{(k)}^2)$. Panel~(b) displays item intercepts $\beta_j^{[s]}$ simulated from $\text{N}(\hat{\beta}_0, \hat{\tau}^2)$. Coverage of the estimated values by the simulated distributions supports successful recovery of the hierarchical structure.
    }
    \label{fig:sim study1}
\end{figure}

\begin{figure}[htb]
    \centering
    \subfloat[Observed responses]{{\includegraphics[width=0.4\textwidth]{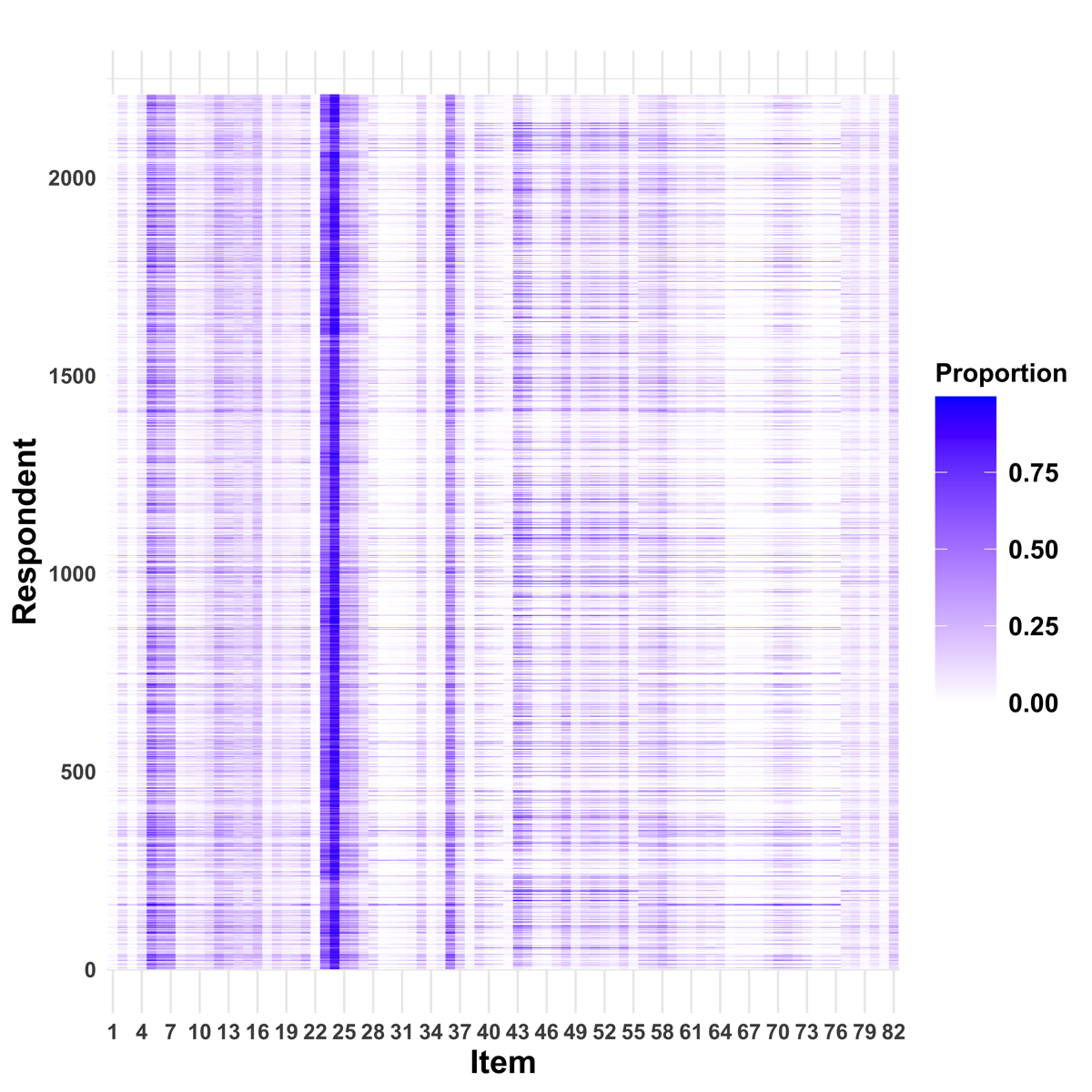}}\label{fig:original data map}}
    \subfloat[Predicted probabilities]{{\includegraphics[width=0.4\textwidth]{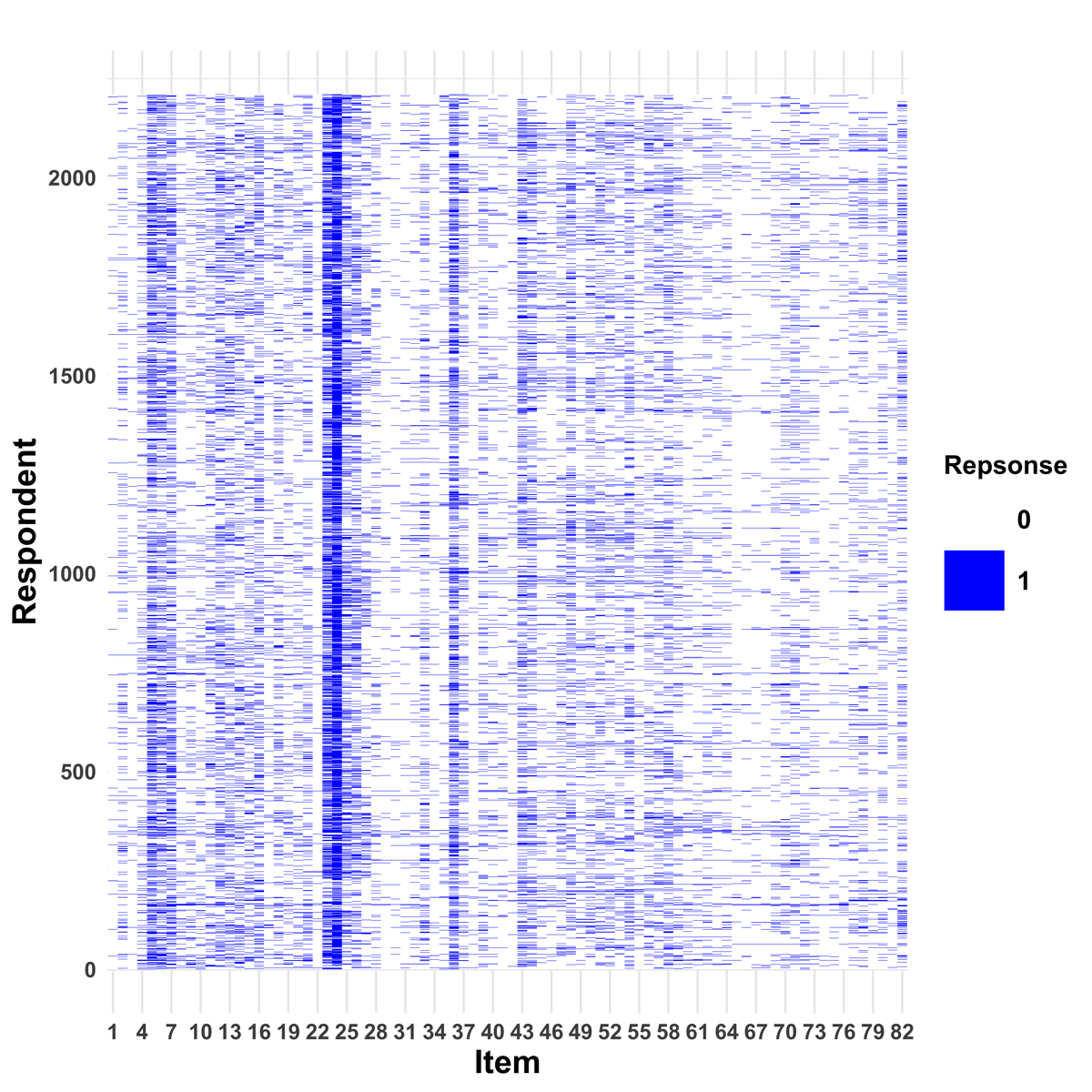}}\label{fig:phat map}}
    \caption{
    Posterior predictive check comparing observed response patterns with model-based predictions. Each cell represents a student-item pair; blue indicates a vulnerability response ($y_{ij(k)} = 1$) in panel~(a) and a predicted probability of vulnerability in panel~(b). Predictions are computed from 5,000 posterior samples.
    }
    \label{fig:ppcheck}
\end{figure}

Figure~\ref{fig:sim study1} shows that the simulated parameter distributions cover the estimated values, confirming that the hierarchical structure is successfully recovered. Figure~\ref{fig:ppcheck} compares observed response patterns with posterior predictive distributions generated from 5,000 posterior samples. The correspondence between panels indicates that HLSIRM adequately reproduces both individual-level response tendencies and school-level patterns.

\subsubsection{Classification Performance}

Table~\ref{table:performance} presents classification performance metrics. The binary classification threshold was selected to maximize the F1 score, which is appropriate given the substantial class imbalance (vulnerability responses comprise 13.1\% of observations). The high specificity (0.911) indicates effective identification of students without mental health vulnerabilities, while the moderate sensitivity (0.618) reflects the inherent difficulty of detecting the minority class in imbalanced data. The AUC of 0.894, which is threshold-independent, demonstrates robust discriminative ability and suggests that the hierarchical structure captures meaningful patterns in the data. In practical implementation, the moderate sensitivity suggests that HLSIRM should be complemented with additional screening mechanisms to ensure identification of all at-risk students.

\begin{table}[hbtp]
\centering
\begin{tabular}{cccccc}
Specificity & Sensitivity & Accuracy & Precision & F1-score & AUC \\ \hline
0.911 & 0.618 & 0.873 & 0.513 & 0.561 & 0.894 \\
\end{tabular}
\caption{
Classification performance metrics for HLSIRM. The binary threshold was selected to maximize the F1 score given class imbalance (13.1\% vulnerability responses). School-level performance metrics are provided in Section~6 of the Supplementary Material, where AUC values range from 0.836 to 0.928 across schools.
}
\label{table:performance}
\end{table}

\subsection{Main Effect Parameters: Overall Vulnerability Patterns}\label{sec:main_effects}

The main effect parameters characterize overall vulnerability tendencies at the school and item levels, providing the baseline for understanding the mental health landscape before considering interaction effects.

\subsubsection{School-Level Vulnerability}

\begin{figure}[hbtp]
    \centering
    \subfloat[$\alpha_{(k)}$]{{\includegraphics[width=0.45\textwidth,height=7cm]{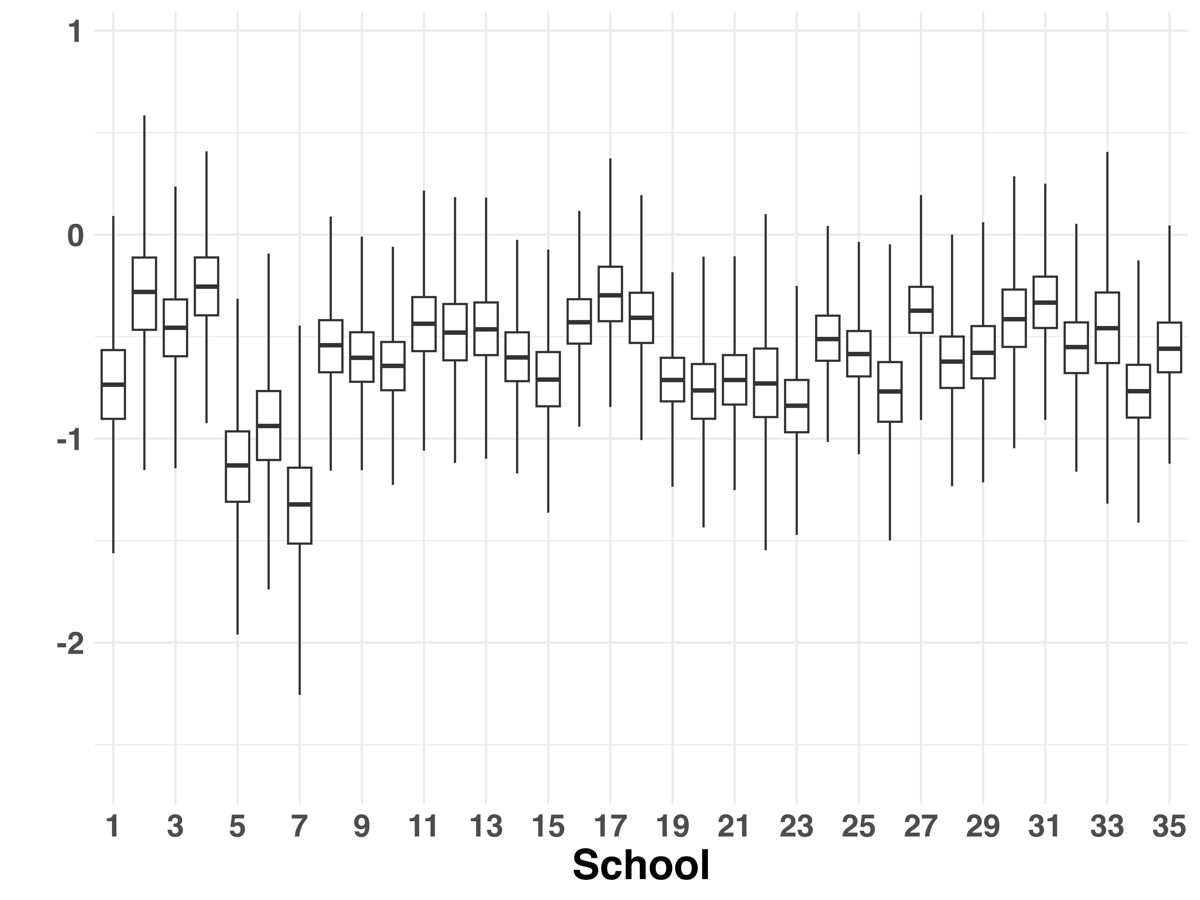}} \label{fig:alphak post samples}}
    \subfloat[$\tilde{\alpha}_{(k)}$]{{\includegraphics[width=0.45\textwidth,height=7cm]{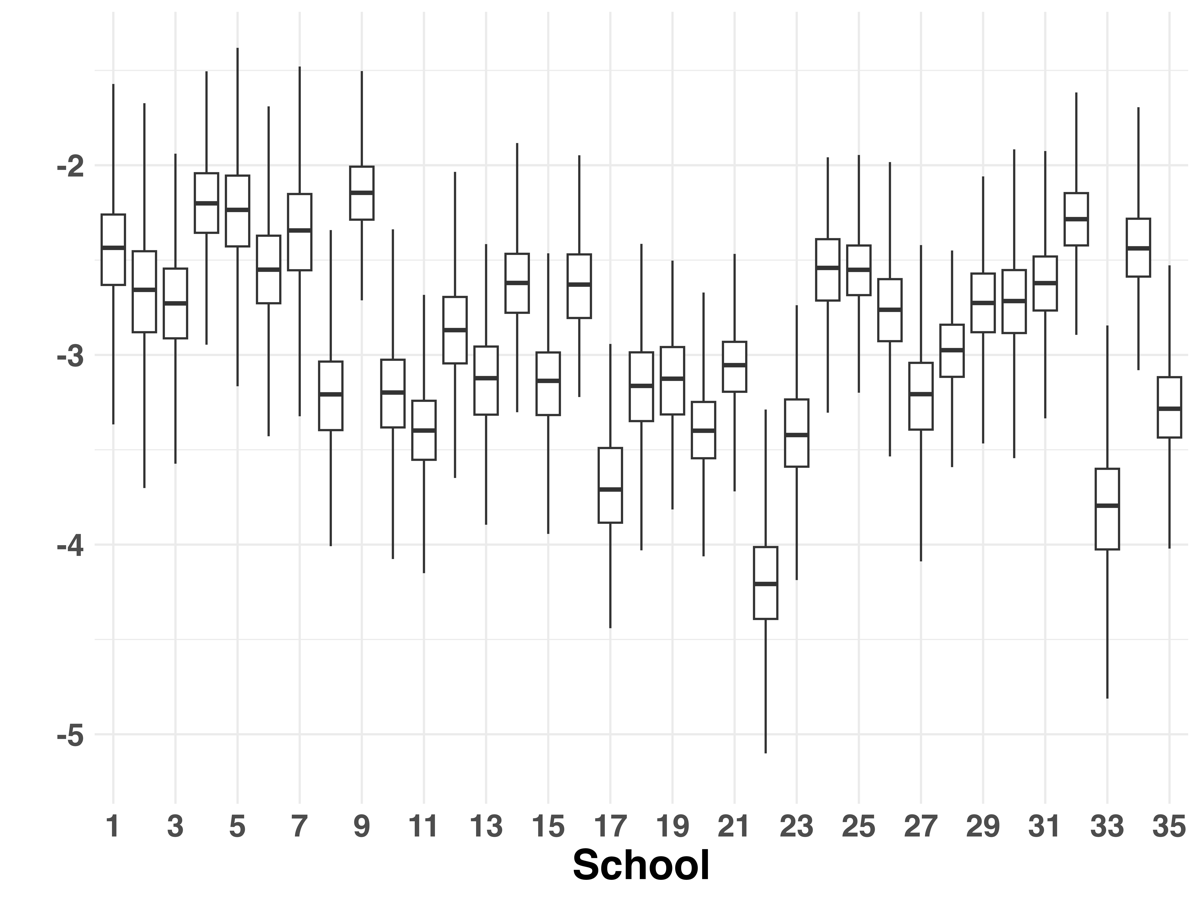}} \label{fig:tilde_alphak}}
    \caption{
    Posterior distributions of school-level vulnerability parameters. Panel~(a) shows the baseline intercept $\alpha_{(k)}$, which captures the average vulnerability of school $k$. Panel~(b) shows the interaction-adjusted vulnerability $\tilde{\alpha}_{(k)} = \alpha_{(k)} + \frac{1}{p}\sum_{j=1}^{p} \mathbf{z}_{(k)}^\top \mathbf{w}_j$, which combines the main effect with the average
    interaction effect across all items. Schools are indexed $1, \ldots, 35$.
    }
\end{figure}

Figure~\ref{fig:alphak post samples} presents the posterior distributions of school-level intercept parameters $\alpha_{(k)}$, representing the average baseline vulnerability of each school's student population; higher values indicate a greater tendency to report mental health vulnerabilities across all items.

The distribution of $\alpha_{(k)}$ across schools centers near zero with slightly negative values, suggesting generally moderate baseline vulnerability levels. However, interpreting school vulnerability solely through $\alpha_{(k)}$ can be misleading because HLSIRM decomposes total vulnerability into main effects and interaction effects. When a school latent position vector aligns closely with the dominant direction of item vectors in the interaction map, the interaction term $\mathbf{z}_{(k)}^\top \mathbf{w}_j$ absorbs a substantial portion of the vulnerability signal for most items, resulting in a relatively lower estimate of $\alpha_{(k)}$. Schools aligned with the dominant vulnerability direction may therefore appear to have low $\alpha_{(k)}$ despite exhibiting substantial vulnerability in specific domains.

To account for this decomposition, we computed the interaction-adjusted measure $\tilde{\alpha}_{(k)} = \alpha_{(k)} + \frac{1}{p}\sum_{j=1}^{p} \mathbf{z}_{(k)}^\top \mathbf{w}_j$, which combines the main effect with the average
interaction effect across all items for school $k$ \citep{kang2024recent}. Because the model operates on the logit scale, $\tilde{\alpha}_{(k)}$ represents the average log-odds of vulnerability responses across the entire item set. Figure~\ref{fig:tilde_alphak} shows that all $\tilde{\alpha}_{(k)}$ values are negative, indicating that the majority of schools demonstrate adequate mental health overall. Despite the elevated stress indicators among Incheon elementary students noted in Section~\ref{sec:introduction}, most schools show generally adequate mental health functioning when aggregated across all domains.

Comparing the two panels reveals that interaction effects substantially influence overall vulnerability assessment. Schools~5 and~7 exhibit the highest $\tilde{\alpha}_{(k)}$ values despite having relatively low $\alpha_{(k)}$, indicating that their elevated vulnerability stems primarily from alignment with specific item clusters rather than generalized mental health difficulties. Conversely, Schools~22 and~33 show the lowest $\tilde{\alpha}_{(k)}$ values due to strongly negative interaction terms, indicating that most students in these schools are positioned opposite to the dominant direction of vulnerability items. School~4 displays elevated values for both $\alpha_{(k)}$ and the average interaction term, suggesting broadly distributed vulnerability that extends beyond the dominant item clusters. School~17 shows a low average interaction term combined with high $\alpha_{(k)}$, suggesting that its vulnerability is concentrated on items outside the main clusters.

\subsubsection{Item-Level Vulnerability}

\begin{figure}[hbtp]
    \centering
    \subfloat[$\beta_j$]{{\includegraphics[width=0.45\textwidth,height=7cm]{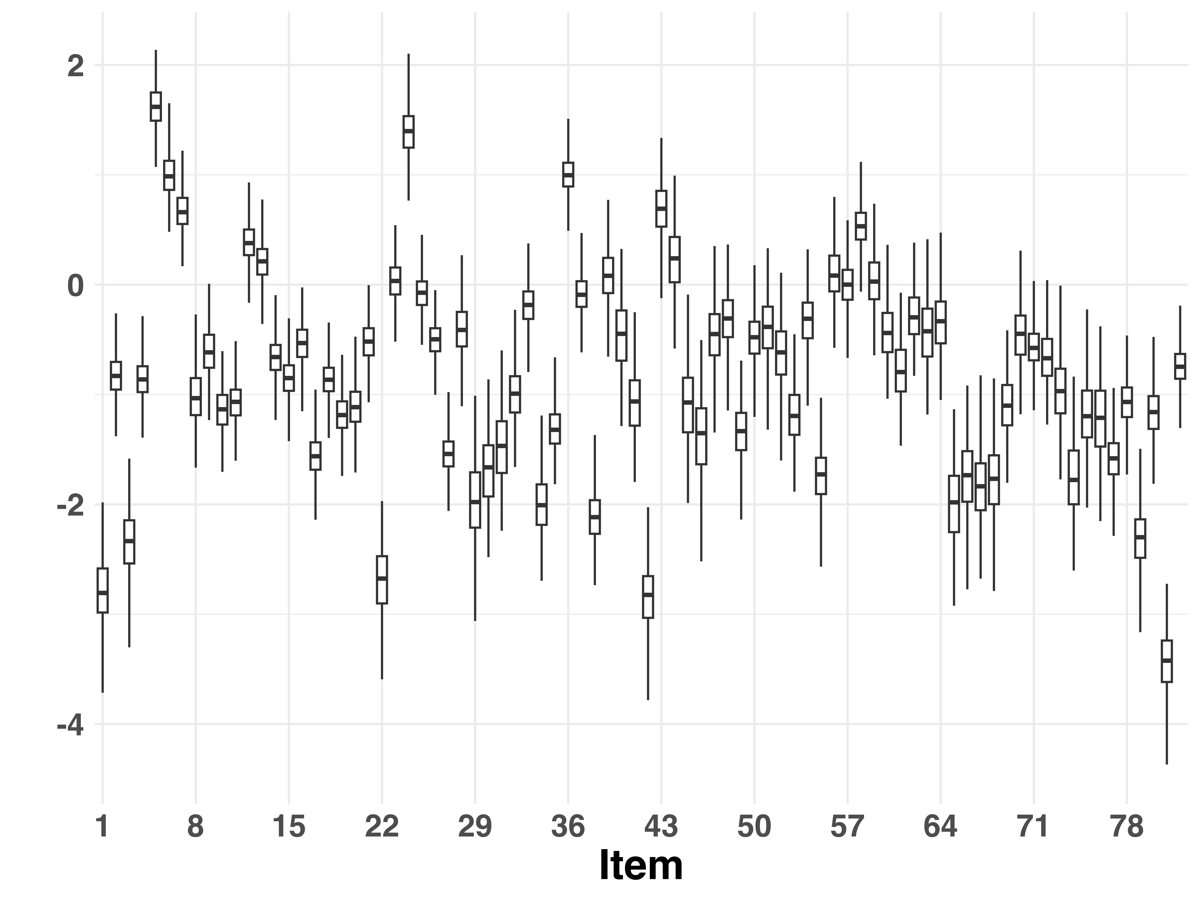}} \label{fig:betaj post samples}}
    \subfloat[$\tilde{\beta}_j$]{{\includegraphics[width=0.45\textwidth,height=7cm]{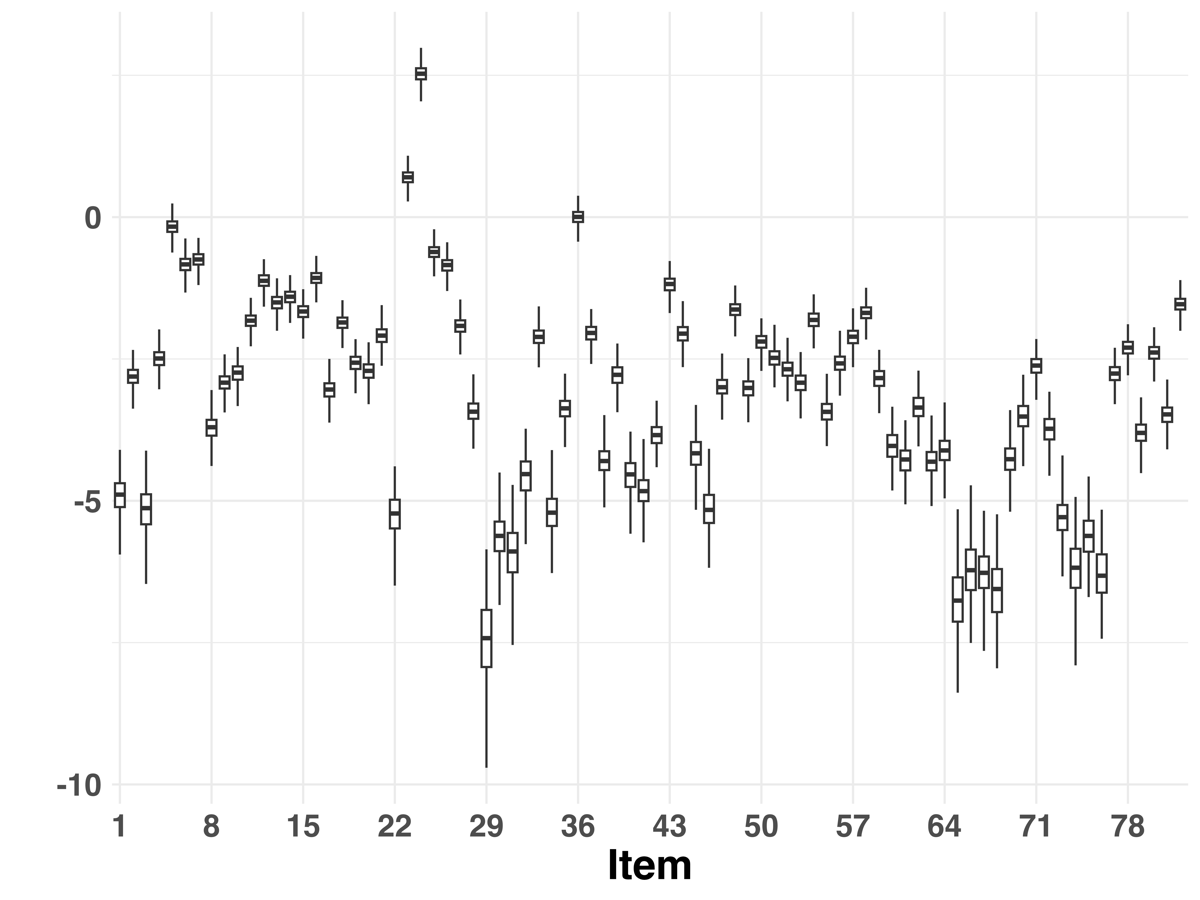}} \label{fig:tilde_betaj}}
    \caption{
    Posterior distributions of item-level vulnerability parameters. Panel~(a) shows the baseline intercept $\beta_j$, which captures the endorsement rate of item $j$. Panel~(b) shows the interaction-adjusted item vulnerability $\tilde{\beta}_j = \beta_j + \frac{1}{K}\sum_{k=1}^{K} \mathbf{z}_{(k)}^\top \mathbf{w}_j$, which combines the main effect with the average
    school-level interaction effect. Items are indexed $1, \ldots, 82$.
    }
\end{figure}

Figure~\ref{fig:betaj post samples} presents the posterior distributions of item intercept parameters $\beta_j$, representing the baseline endorsement rate of each item across all respondents; higher values indicate items more frequently endorsed as vulnerabilities. The $\beta_j$ values span both negative and positive ranges, with substantial heterogeneity across items. To account for interaction effects, we computed $\tilde{\beta}_j = \beta_j + \frac{1}{K}\sum_{k=1}^{K} \mathbf{z}_{(k)}^\top \mathbf{w}_j$, which represents the average log-odds of vulnerability responses for each item across all schools \citep{kang2024recent}.

Figure~\ref{fig:tilde_betaj} reveals that only Items~23 and~24, which assess the absence of counseling experience, exhibit positive $\tilde{\beta}_j$ values, indicating substantial vulnerability. Items~5, 6, and~36, which showed positive $\beta_j$ values, demonstrate negative interaction effects with school positions, meaning they function as non-vulnerable items when actual response probabilities are computed. This pattern highlights the importance of considering interaction effects: assessing item vulnerability based solely on $\beta_j$ would overestimate the prevalence of certain difficulties.

Items with the most negative $\tilde{\beta}_j$ values include Items~29 (school belonging), 65--68 (understanding others), and 73--76 (collaborative problem-solving). For these items, negative interaction effects outweigh their baseline parameters, indicating that at the average school, students demonstrate adequate competencies in school belonging and interpersonal understanding.

\subsubsection{Individual Student Vulnerability}

\begin{figure}[hbtp]
    \centering
    \includegraphics[width=0.6\textwidth,height=7.5cm]{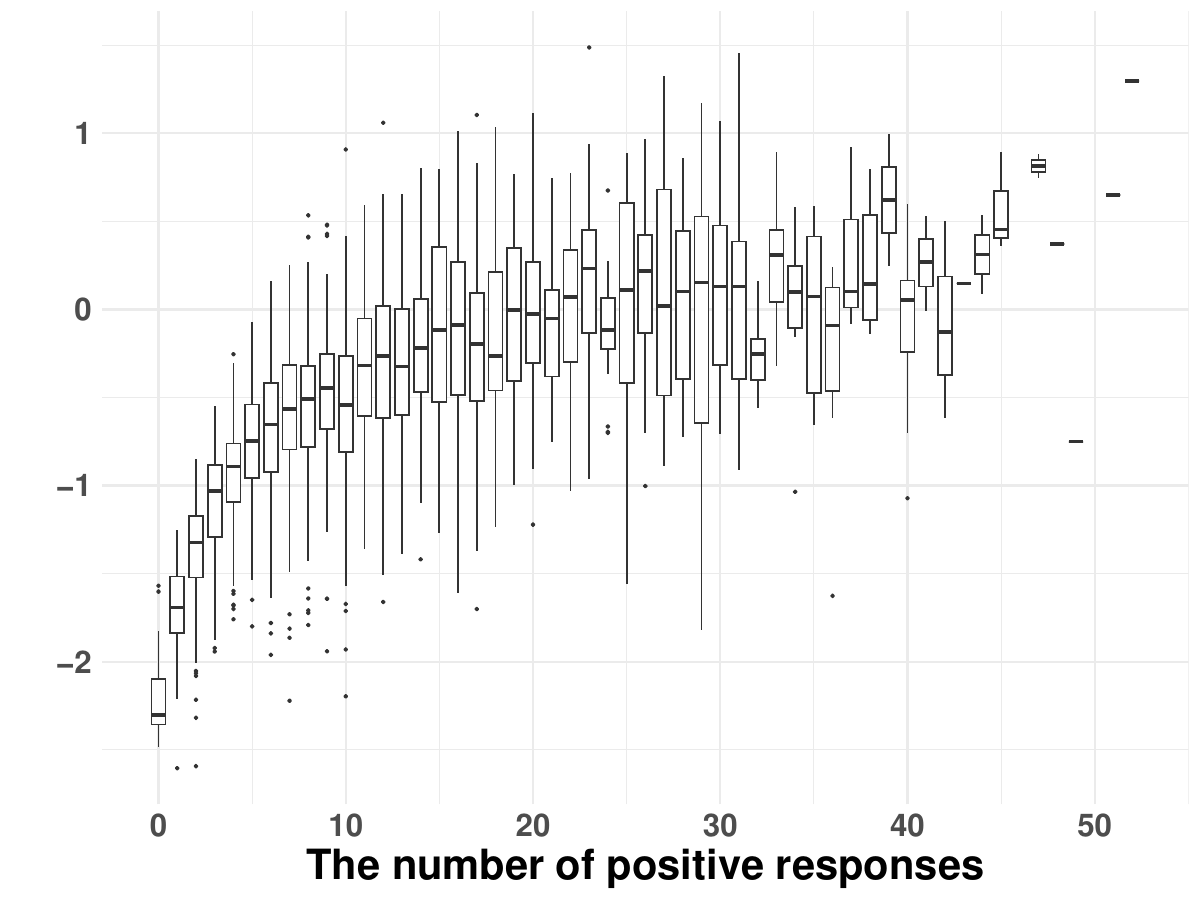}
    \caption{
    Posterior distributions of individual intercepts $\alpha_{i(k)}$, grouped by the number of vulnerability responses endorsed. The positive monotonic relationship indicates that $\alpha_{i(k)}$ captures individual vulnerability differences. 
    }
    \label{fig:alphaki}
\end{figure}

Figure~\ref{fig:alphaki} presents the relationship between estimated individual intercepts $\alpha_{i(k)}$ and the number of vulnerability responses endorsed. The positive association indicates that HLSIRM captures individual differences in mental health vulnerability: students who endorse more items as vulnerabilities receive higher $\alpha_{i(k)}$ estimates.

The relationship shows an initial monotonic increase followed by a plateau at higher endorsement frequencies. This pattern occurs because students with many vulnerability responses increasingly align with the corresponding item clusters in the interaction map, shifting explanatory power from the main effect to the interaction term. The reduced variance at higher frequencies reflects data sparsity and Bayesian shrinkage, as fewer students occupy the extreme vulnerability range.

\subsection{Interaction Map: Item Clustering and School Positioning}\label{sec:interaction_map}

The main effect analysis revealed that interaction effects substantially influence vulnerability assessment at both the school and item levels. The interaction map constitutes the core analytical output of HLSIRM, enabling visualization of relationships among items and schools that reveal domain-specific vulnerability patterns. We first examine item clustering to identify coherent domains of mental health vulnerability, then analyze school positioning to understand how different schools relate to these domains.

\subsubsection{Item Position Vectors and Clustering}

\begin{figure}[hbt]
    \centering
    \subfloat[Item position vectors $\mathbf{w}_j$]{{\includegraphics[width=0.45\textwidth, height=7cm]{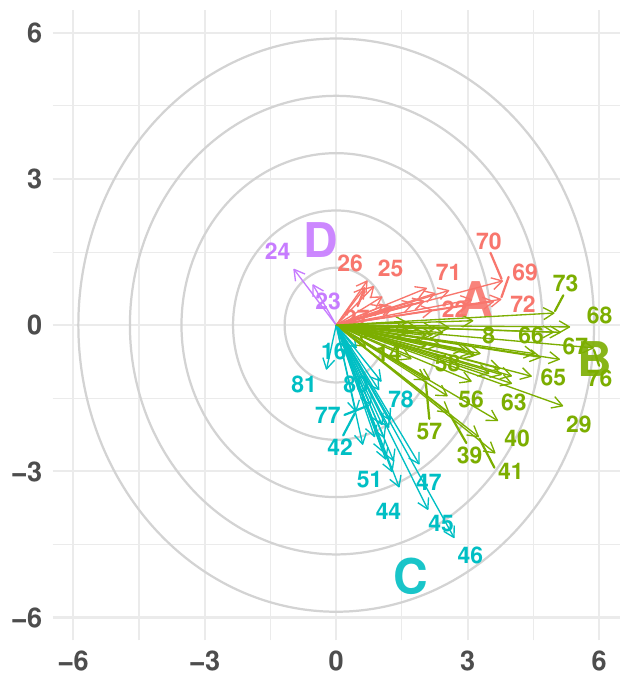}} \label{fig:vj map}}
    \subfloat[Inner product heatmap]{{\includegraphics[width=0.45\textwidth,height=7cm]{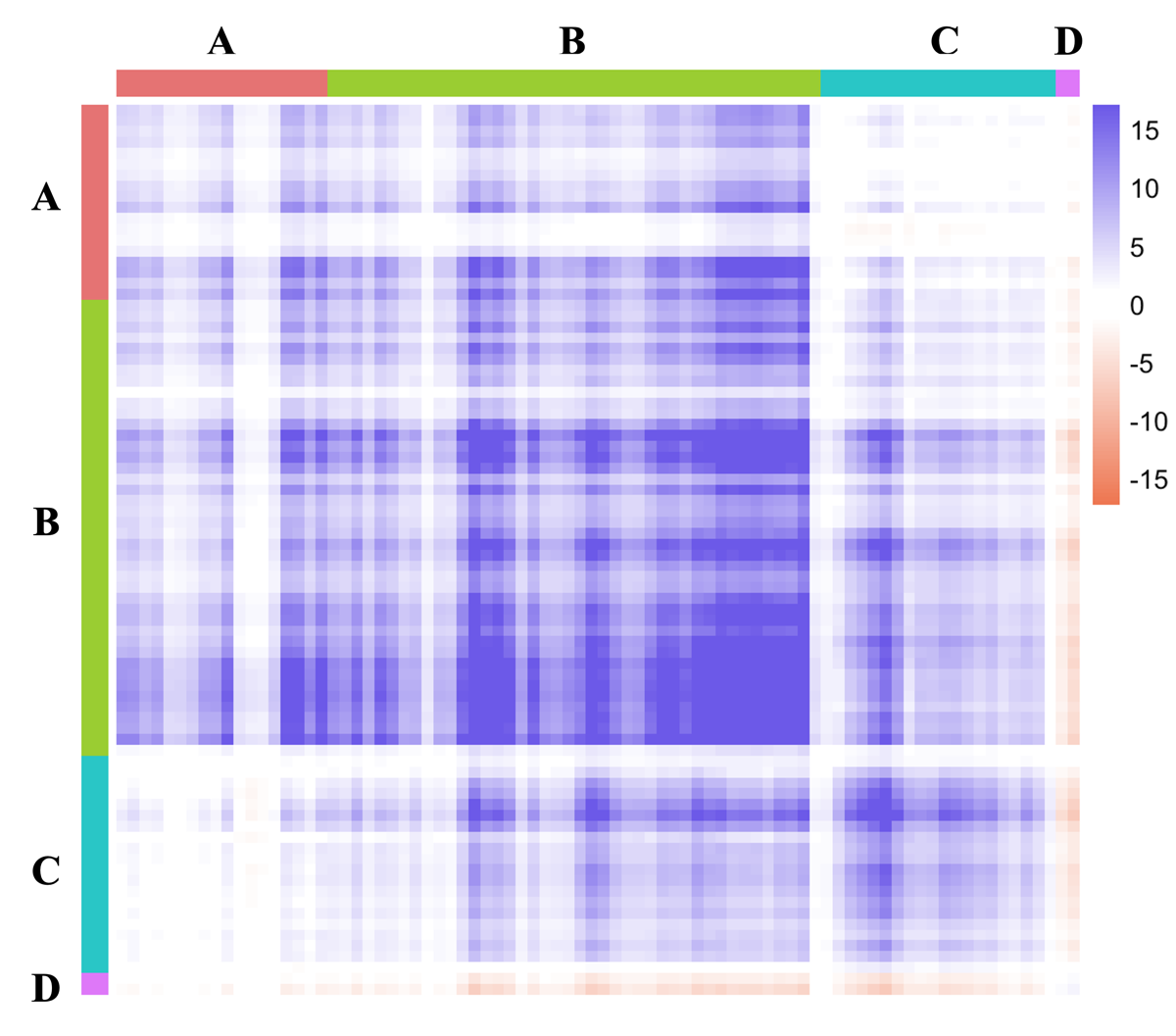}} \label{fig:vj tile map}}
    \caption{
    Item latent vectors in the interaction map. Panel~(a) shows the posterior mean of each item position vector $\mathbf{w}_j$, with colors and labels indicating clustering assignments (four clusters: A--D). Vector direction indicates the type of student-item interaction and vector magnitude indicates discrimination strength. Panel~(b) shows the matrix of pairwise inner products $\mathbf{w}_j^\top \mathbf{w}_{j'}$ between all item vectors: blue indicates positive interaction (items co-endorsed), red indicates negative interaction (items inversely related), and white indicates independence.
    }
\end{figure}

Figure \ref{fig:vj map} presents the item interaction map, where arrows represent latent position vectors with numerical labels corresponding to item indices. Vector direction indicates the type of interaction (similar directions imply positive association), while vector magnitude indicates the strength of discrimination among students.

To identify coherent vulnerability domains, we applied spectral clustering to the item position vectors using cosine similarity \citep{racolte2022spherical}. The number of clusters was selected by evaluating silhouette coefficients and Davies-Bouldin indices across candidate values. At $k = 4$, the silhouette coefficient \citep{rousseeuw1987silhouettes} reached its maximum (0.76) and the Davies-Bouldin index \citep[DBI;][]{davies2009cluster} its minimum (0.36), indicating strong between-cluster separation and within-cluster cohesion. Table~\ref{table:clustering result} presents the resulting cluster assignments. Item cluster maps for $k = 2, \ldots, 7$ are provided in Section~7 of the Supplementary Material.

\begin{table}[hbtp]
\centering
\begin{tabular}{cll}
Group & Item index & Item category \\ \hline
\textbf{A} & 5-7, 10; 11,15,18,20-22; 25-27; 36; 69-72; & SH; DH; SIV; SAD; SS \\
\textbf{B} & 1-4,8,9; 12-14,17,19; 28-35,37,38; 39-41; 56-64; 65-68, 73-76; 82; & SH; DH; SAD; HP; EM; SS; DGT \\
\textbf{C} & 16; 42-47; 48-55; 77-80; 81; & DH; FD; ST; SPD; DGT \\
\textbf{D} & 23-24 & CE \\ 
\end{tabular}
\caption{
Clustering results for item position vectors ($k = 4$, silhouette coefficient $= 0.76$, DBI $= 0.36$). Items cluster by empirical response patterns rather than predefined survey categories. Item indices correspond to those defined in Table~\ref{Tab:Table2}.
}
\label{table:clustering result}
\end{table}

The clustering results reveal several insights about the structure of mental health vulnerability among elementary students, some of which diverge from the predefined survey categories.
\begin{itemize}
\item {\bf Cluster A: Disengagement and Non-Participation.} This cluster encompasses items related to low classroom community spirit (Items~69--72), reduced engagement in school activities (Items~25--27), diminished importance placed on healthy lifestyle habits, and inadequate study planning. Items measuring low classroom participation exhibit substantial vector magnitudes and spatial proximity, suggesting strong mutual association: students endorsing Item~70 (``I do not actively participate in resolving classroom problems'') are more likely to also endorse Item~72 (``I do not proactively show warmth to isolated classmates''). Item~36 from the School Adaptation category (``I have no teacher I would want to consult when troubled'') belongs uniquely to Cluster~A rather than clustering with other SAD items, suggesting that unwillingness to seek teacher support relates more closely to disengagement patterns than to general school adaptation difficulties.

\item {\bf Cluster B: Psychosocial Difficulties.} This cluster contains the majority of items reflecting compromised mental health functioning or relational vulnerability, including unhappiness indicators (Items~39--41), poor emotional management (Items~56--64), low school belonging and weak peer relationships (Items~28--35, 37--38), and interpersonal deficits (Items~65--68, 73--76). The co-clustering of these domains suggests that students experiencing difficulties with emotional regulation and school belonging are more likely to report compromised well-being overall. Within the cluster, items concerning communication and collaborative difficulties (Items~65--68, 73--76) align toward Cluster~A, unhappiness indicators (Items~39--41) align toward Cluster~C, and impaired school adaptation and emotional management items (Items~28--35, 37, 38, 56--64) occupy intermediate positions, reflecting a gradient of mental health vulnerabilities from social functioning to affective states.

\item {\bf Cluster C: Stress, Depression, and Digital Coping.} This cluster captures associations among stress-related indicators (Items~48--55), anxiety and depression (Items~42--47), and smartphone dependency (Items~77--80). The spatial proximity suggests that stress among elementary students co-occurs with emotional anxiety and that smartphone use may function as a coping mechanism. The close positioning of Item~46 (``I feel sad and depressed''), Item~52 (``I feel anxious about peer relationships''), Item~79 (``I feel alone without my smartphone''), and Item~50 (``Homework is too much or too difficult'') points to an interconnected cycle of academic pressure, emotional distress, and digital dependency that warrants integrated intervention approaches.

\item {\bf Cluster D: Counseling Experience.} This cluster contains only Items~23 and~24, assessing the absence of counseling experience with homeroom teachers and professional counselors, respectively. The close positioning of these two items indicates that elementary students treat different counseling resources similarly rather than differentiating between them, suggesting a pattern of either seeking help through multiple channels or avoiding counseling altogether.
\end{itemize}

\paragraph*{Cluster Configuration and Independence.} Figure~\ref{fig:vj tile map} visualizes the pairwise inner products between item vectors, revealing both the strength of interactions and the independence structure among clusters. Cluster~A forms nearly perpendicular orientations to Clusters~C and~D, indicating that disengagement-related vulnerabilities and stress-depression vulnerabilities operate through distinct mechanisms. This orthogonality has a practical implication: interventions targeting one domain are unlikely to adversely affect the other, enabling schools to address specific vulnerability areas without concern for spillover effects.

The clustering results also show that items from the same predefined category can separate based on empirical response patterns. Study Habits items with high vulnerability response rates (Items~5--6) appear in Cluster~A, while those with lower rates (Items~1, 3) aggregate in Cluster~B. Similarly, Anxiety and Depression items measuring negative emotional experiences (Items~42--47) cluster in~C, while Emotional Management items assessing regulation in challenging situations (Items~56--58) appear in~B. This separation confirms that students perceive emotional distress and regulatory competencies as distinct domains rather than attributing negative emotions to regulatory deficiencies.

\subsubsection{School Position Vectors and Regional Patterns}

\begin{figure}[hbt]
    \centering
    {\includegraphics[width=0.5\textwidth]{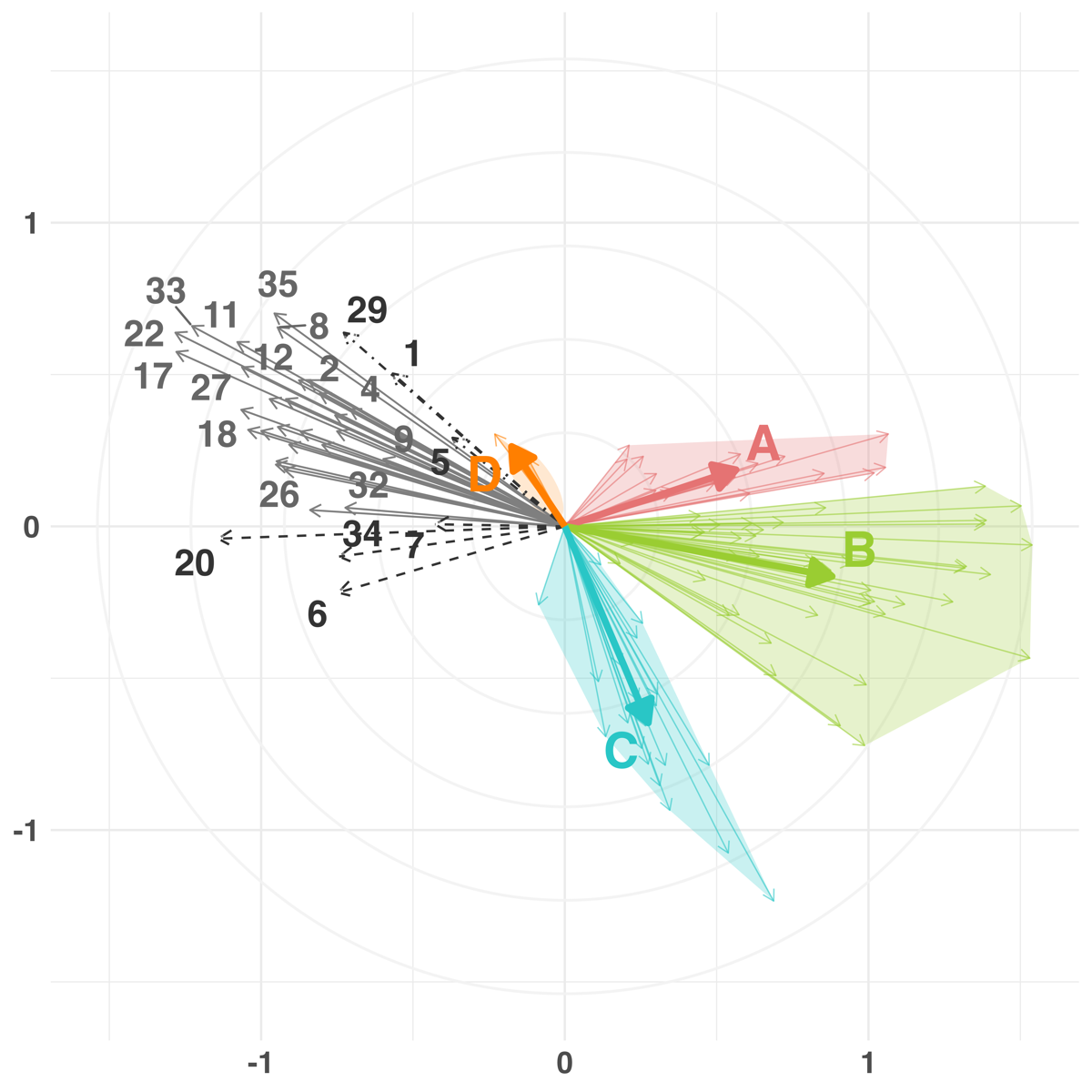}}
    \caption{
    School and item latent vectors in the interaction map. School vectors $\mathbf{z}_{(k)}$ (black) and item vectors $\mathbf{w}_j$ (colored by cluster assignment from Table~\ref{table:clustering result}) are plotted together; vector alignment indicates the direction and strength of interaction under the inner product formulation. School vectors are differentiated by line style: dotted lines indicate schools proximate to Cluster~A, dashed lines indicate schools proximate to Cluster~C, and solid lines indicate remaining schools. Item vectors are displayed at reduced scale relative to Figure~\ref{fig:vj map} to emphasize angular relationships.
    }
    \label{fig:uk_map}
\end{figure}

Figure~\ref{fig:uk_map} presents the school position vectors (dark gray) alongside item vectors in the interaction map. Each school vector represents the aggregate response pattern of students within the corresponding institution, enabling direct identification of school-specific vulnerability domains through angular relationships with item vectors. Item vectors in Figure~\ref{fig:uk_map} are displayed at reduced scale to facilitate visual comparison; actual item vector magnitudes are shown in Figure~\ref{fig:vj map}.

A notable pattern emerges from the overall positioning: with the exception of Cluster~D (counseling experience), most item vectors point in directions opposite to the main concentration of school vectors. This configuration indicates that students across most
schools demonstrate adequate functioning rather than vulnerability in the assessed mental health domains. The pattern provides important context for interpreting the school-specific results that follow: while certain schools exhibit elevated vulnerability in particular domains, the overall picture reflects generally adequate mental health among Incheon elementary students.

\paragraph*{Counseling Experience: The Primary Vulnerability Domain.} 
The items most closely aligned with school vectors are those assessing the absence of counseling experience (Cluster~D). However, interpreting this alignment requires consideration of school context rather than attribution to student-level psychological barriers. Most schools positioned closest to Items~23 and~24 lacked professional school counselors at the time of data collection. Schools~1 and~2 in Ongjin-gun (rural island area), for example, have no on-campus counseling facilities, and the considerable physical distance to government-affiliated counseling centers poses a significant geographical barrier to service utilization. The absence of counseling experience thus cannot be attributed solely to students' reluctance to seek help but likely reflects inadequate infrastructure.

The counseling items are positioned opposite to Cluster~C (stress, depression, smartphone dependency). This spatial arrangement suggests that students with lower stress and depression levels may not perceive a need for counseling, potentially reflecting adequate psychological states rather than unmet needs. Lack of counseling experience may therefore indicate inadequate access, low perceived need due to good mental health, or psychological barriers to help-seeking; school-specific investigation is required to distinguish among these possibilities.

Schools~6, 7, 20, and~34, which are positioned furthest from the counseling items, are also located opposite to Items~35--38, which capture discomfort in teacher--student relationships. In particular, these schools are farthest from Item~36 (``There is no teacher I wish to consult with about my concerns''). This spatial arrangement suggests that students who perceive teachers as attentive and approachable may exhibit less resistance to seeking counseling. Implementation records from the survey year confirm that these schools conducted active internal and external counseling programs \citep{consult2024}.

\paragraph*{Stress and Depression Vulnerabilities.}
Schools~6 and~34 (dashed vectors in Figure~\ref{fig:uk_map}) form acute angles with Cluster~C, indicating strong interactions with stress, depression, and smartphone dependency items. The most proximate items are Item~48 (``I experience stress from grades or exams''), Item~42 (``I have no interest or enthusiasm in anything''), and Item~81 (extended academic device usage). These results corroborate the academic stress patterns described in Section~\ref{sec:introduction} and suggest associations with apathy and diminished interest. The concentration of these vulnerabilities in specific schools rather than across all schools suggests that targeted interventions at these institutions could address a substantial portion of stress-related mental health challenges.

\paragraph*{Disengagement Vulnerabilities.}
Schools~1, 5, and~29 (dotted vectors in Figure~\ref{fig:uk_map}) demonstrate strong interactions with Items~25, 26, and~27 (Cluster~A), which measure low participation in school activities such as meetings, clubs, and events. The positioning of these schools suggests systematic disengagement patterns that may warrant attention to school climate and participatory culture rather than individual-level interventions.

% \paragraph{Regional Differentiation.} 

\subsubsection{School Differences by School Characteristics}

\begin{figure}[hbt]
\centering
\subfloat[Regional type]{{\includegraphics[width=0.33\textwidth]{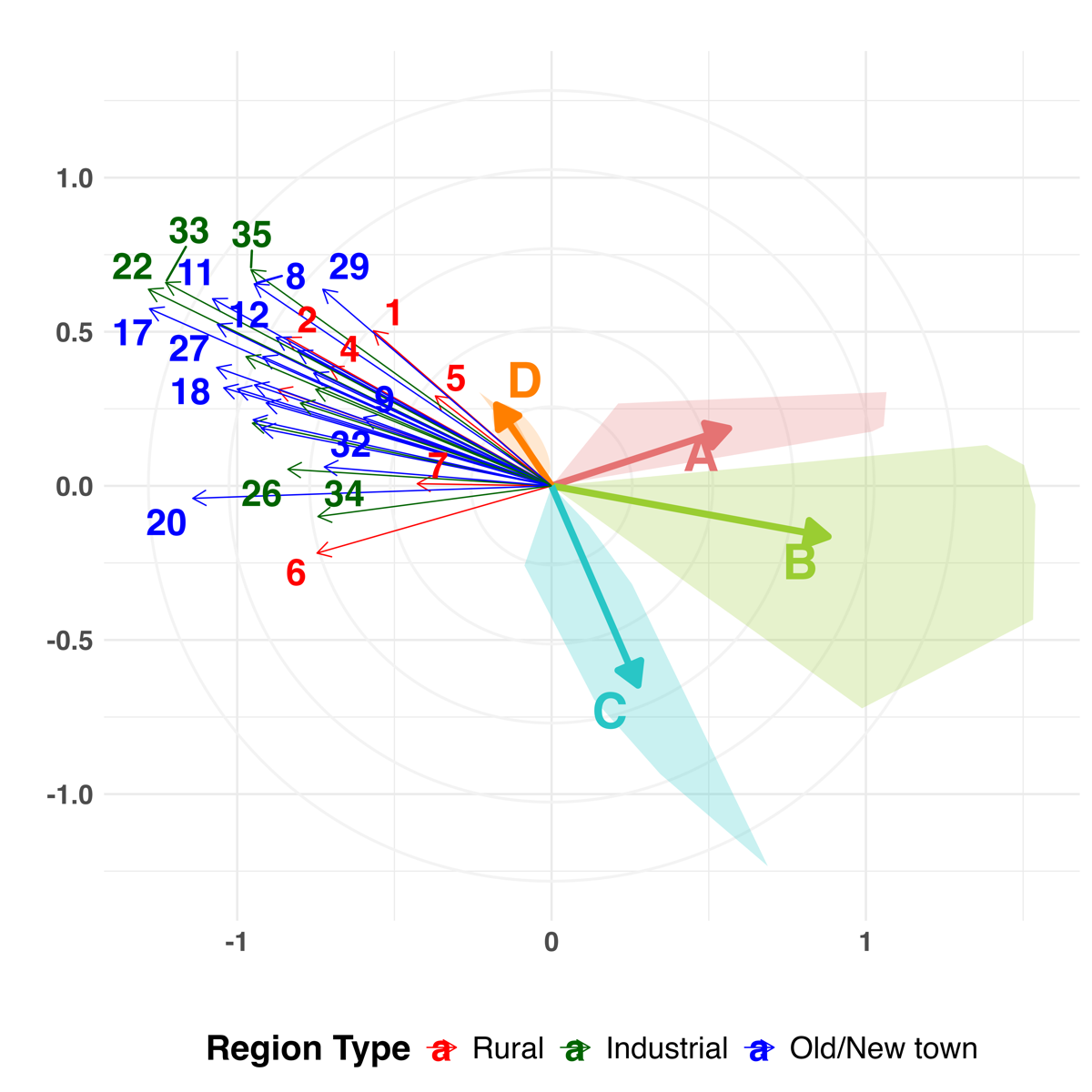}}\label{fig:region}}
\subfloat[Bullying prevention hours]{{\includegraphics[width=0.33\textwidth]{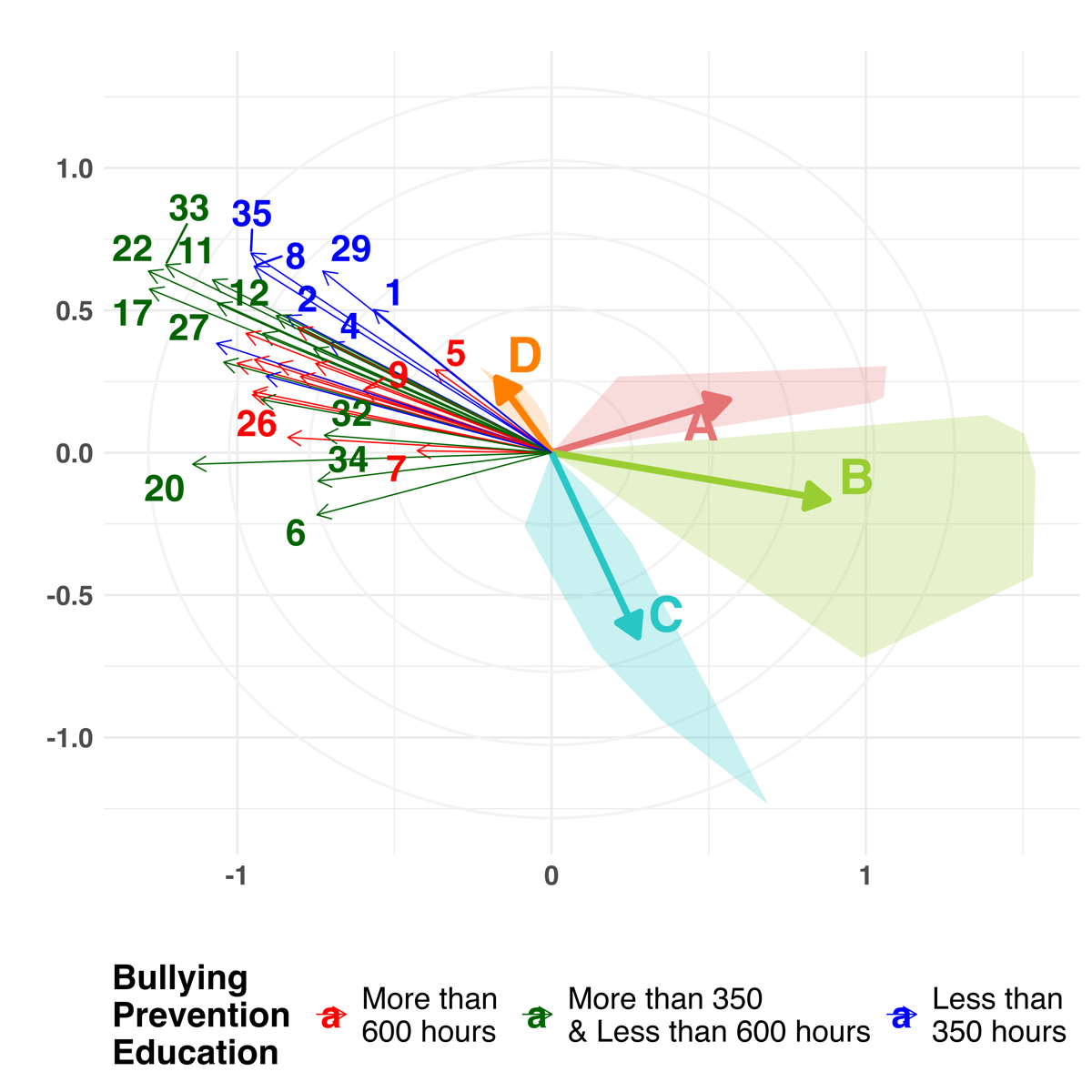}}\label{fig:bully}}
\subfloat[Safety education hours]{{\includegraphics[width=0.33\textwidth]{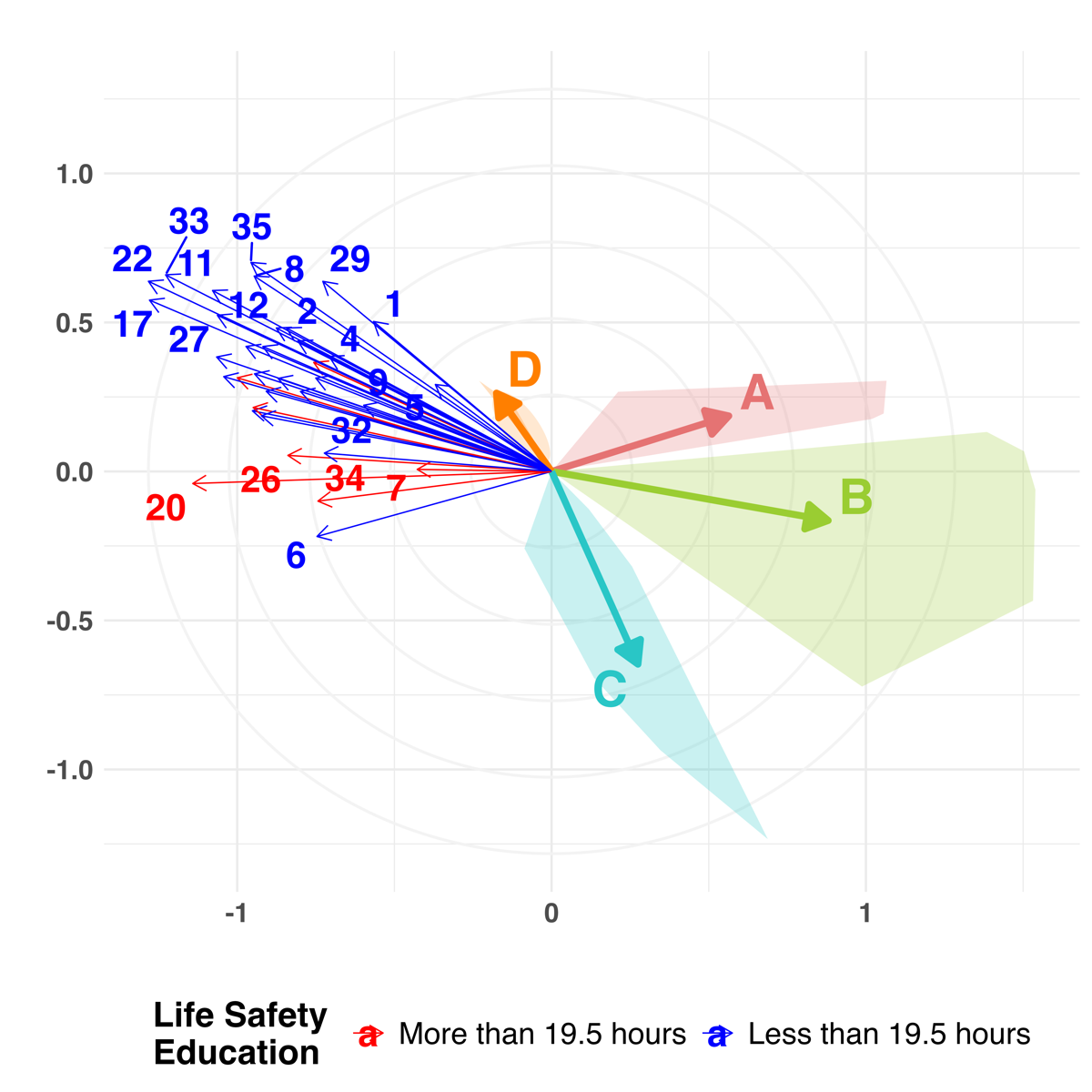}}\label{fig:safe}}
\caption{
School position vectors $\mathbf{z}_{(k)}$ in the interaction map, colored by (a)~regional classification, (b)~annual school bullying prevention education hours, and (c)~annual daily life safety education hours. Colors in each panel correspond to the
respective legends; school labels indicate school indices ($1, \ldots, 35$). Item vectors are omitted to reduce visual clutter; their positions are shown in Figures~\ref{fig:vj map} and~\ref{fig:uk_map}.
}
\end{figure}

Figure~\ref{fig:region} illustrates school positions colored by regional type, enabling examination of whether mental health patterns systematically vary across the socioeconomic contexts of Incheon. Rural island schools (Schools~1, 5, 6, 7; red in Figure~\ref{fig:region}) occupy predominantly peripheral positions, exhibiting response patterns distinct from urban schools. This peripheral positioning likely reflects the challenges of geographically isolated communities, including limited access to mental health resources, smaller peer networks, and different family structures. Industrial area schools (Schools~22, 33, 35; green in Figure~\ref{fig:region}) exhibit the largest vector magnitudes among regional classifications, indicating relatively strong and coherent response tendencies. This pattern may reflect the concentrated socioeconomic characteristics of these districts, where working-class family compositions and economic pressures create more homogeneous student experiences compared to other areas.

Figures~\ref{fig:bully} and~\ref{fig:safe} examine whether school-level educational programs show expected associations with student mental health patterns, providing a form of construct validation for the HLSIRM results. Schools with more hours of bullying prevention education (red and green in Figure~\ref{fig:bully}) predominantly align opposite to Cluster~B items concerning peer relationships, particularly Item~31 (``I do not get along well with my friends'') and Item~34 (``I have no friends to play with''). This positioning indicates stronger competencies in sustaining friendships among students at these schools. The correspondence between program intensity and peer relationship outcomes aligns with established research on bullying prevention effectiveness \citep{farrington2002effectiveness, ehiri2017primary} and suggests that school positions in the interaction map reflect genuine institutional influences on student mental health.
%\textcolor{red}{1. these school-level covariates should be discussed earlier in the data section;}
%

%\textcolor{red}{2. maybe contrast red schools vs other schools, since we discuss red schools only;}
% Figure 11(a)의 지역 라벨은 Rural외에도, Industrial을 포함합니다. 그리고 Figure 11(b)도 red+green이 item Cluster B와 반대 방향으로 향하여 의미있다고 생각하였습니다. 따라서, 지역 라벨 중 언급이 안되는 Old downtown과 Newtown을 Old/New town으로 통합하여 최대 3가지 색상으로 구분하도록 수정했습니다.

%\textcolor{red}{3. add school colors in the text. same for the next paragraph}
% 설명시 추가하였습니다. 

Similarly, schools implementing extensive daily life safety education programs (red in Figure~\ref{fig:safe}) are positioned opposite to Items~12 and~17--22, which assess unhealthy lifestyle habits and health awareness. These items measure insufficient awareness of healthy living and negative daily habits---content that directly corresponds to what safety education programs address. Schools with fewer safety education hours are positioned closer to Cluster~A, which encompasses low health awareness and insufficient self-regulation.

%These associations enhance confidence in the validity of HLSIRM results by demonstrating expected relationships between school-level programs and student response patterns. While causal interpretation requires caution, the alignment between program content and student outcomes suggests that the identified school-level patterns reflect meaningful institutional characteristics.

%Figure \ref{fig:student} presents results categorized using the mean enrollment of 725 students across 35 schools collected in 2024 as the reference point. Considering the direction of the non-counseling experience cluster {\bf D}, no substantial association with infrastructure differences based on school size was observed. 

\subsection{School-Specific Respondent Latent Positions}\label{sec:intervention}

\begin{figure}[htb]
    \centering
    \subfloat[School 35: $\mathbf{z}_{i(35)}$]{{\includegraphics[width=0.33\textwidth]{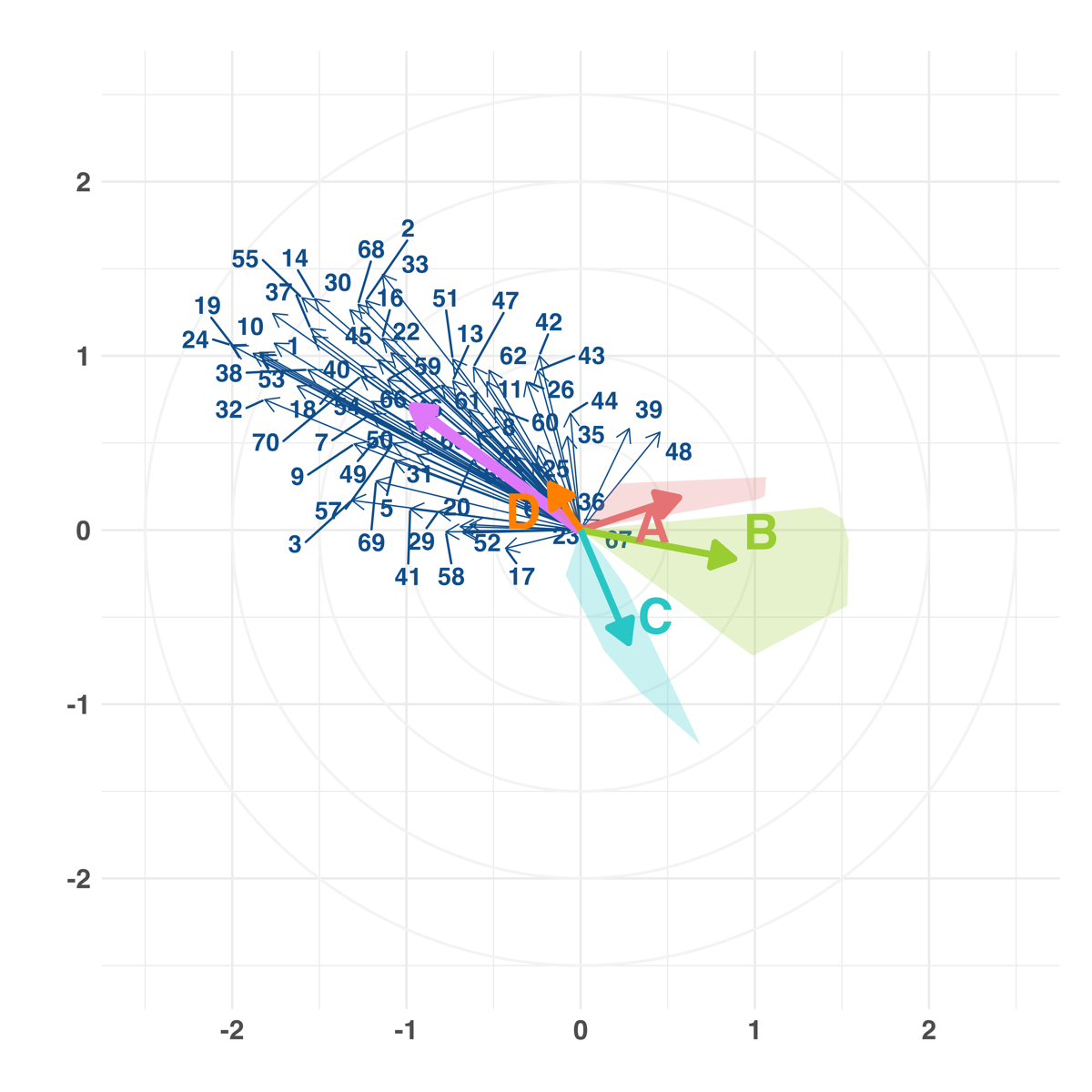}}\label{fig:u35}}
    \subfloat[School 6: $\mathbf{z}_{i(6)}$]
    {{\includegraphics[width=0.33\textwidth]{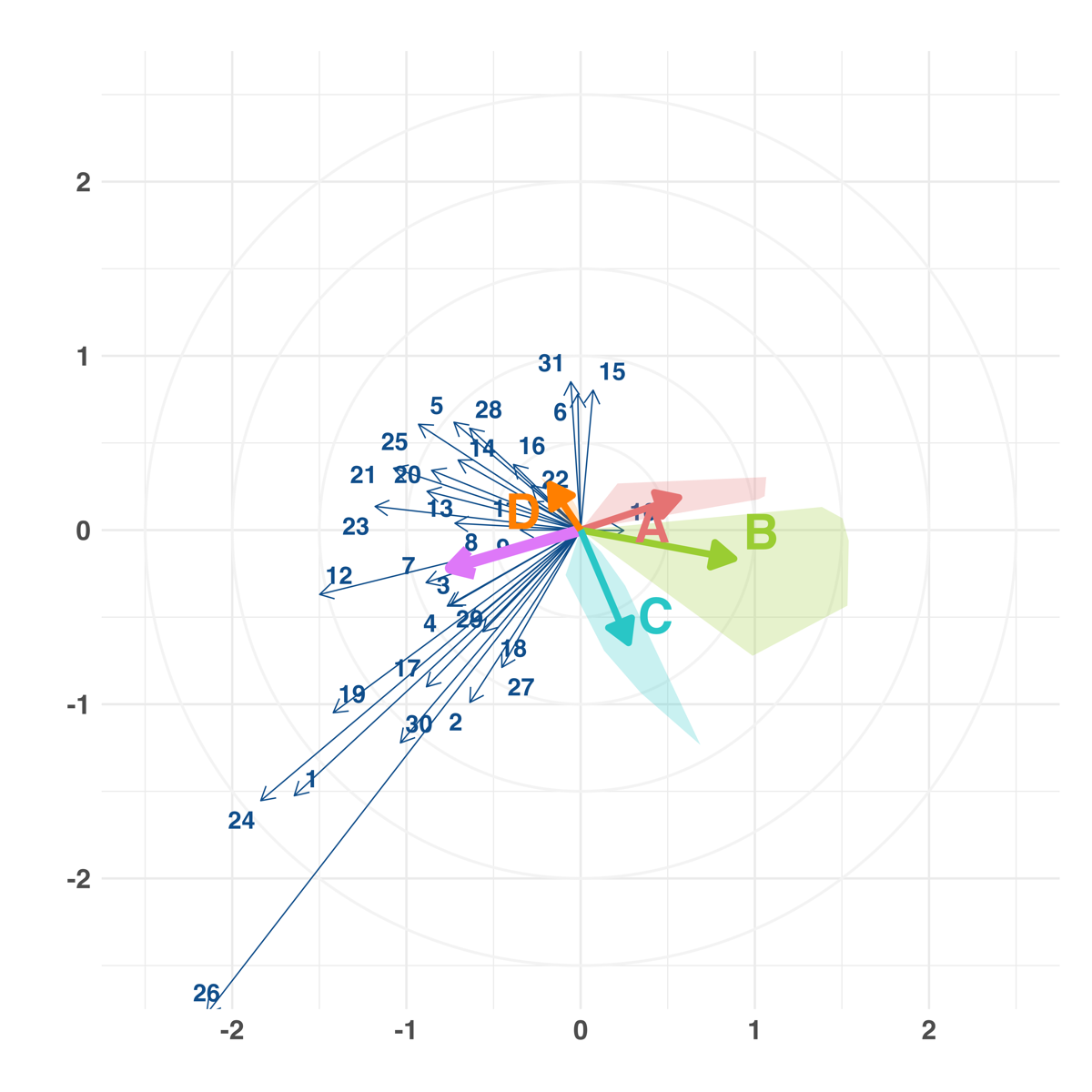}}\label{fig:u06}}
    \subfloat[School 9: $\mathbf{z}_{i(9)}$]{{\includegraphics[width=0.33\textwidth]{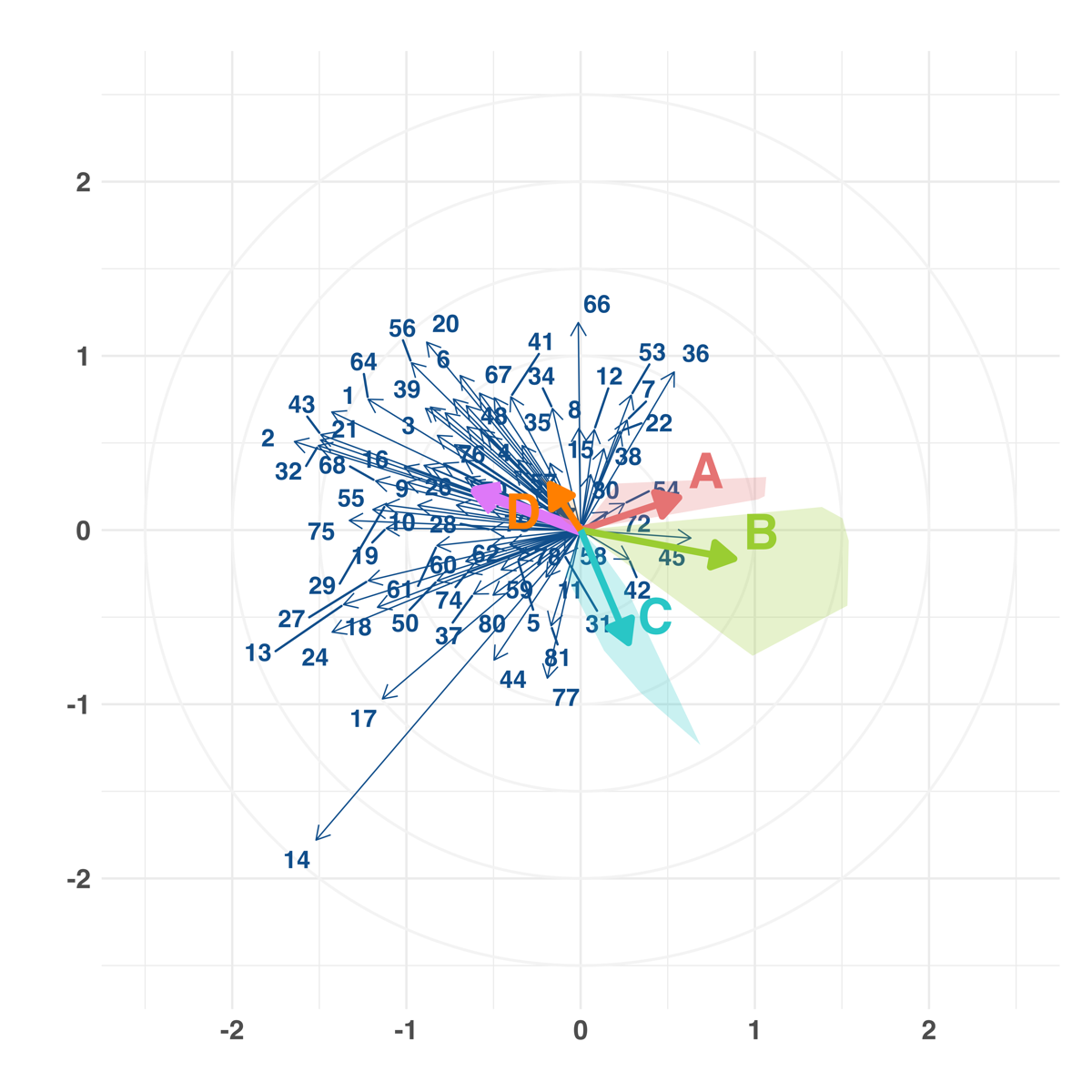}}\label{fig:u09}}
    \caption{
    Individual student position vectors $\mathbf{z}_{i(k)}$ for three schools with distinct vulnerability profiles, plotted on a common scale ($-2.5$ to $2.5$). Panel~(a): School~35 (generally stable mental health); panel~(b): School~6 (stress and depression vulnerabilities); panel~(c): School~9 (heterogeneous student patterns). The school-level latent vector $\mathbf{z}_{(k)}$ is shown in purple in each panel. Item cluster centers from Figure~\ref{fig:vj map} are indicated for reference.
    }
    \label{fig:resp}
\end{figure}

The primary practical contribution of HLSIRM lies in its ability to identify school-specific intervention targets by examining both aggregate school positions and within-school student distributions. This subsection demonstrates how the model outputs translate into actionable insights for three schools representing distinct vulnerability profiles.

\paragraph*{School 35: Generally Stable Mental Health.}
Most student vectors in School~35 align opposite to the centers of item Clusters~B and~C with large magnitudes, contributing to a correspondingly large school vector in the same direction. The resulting negative interactions or independence with Clusters~A, B, and~C indicate generally stable mental health across the student body (see Section~8 of the Supplementary Material for individual-level details). The principal exceptions are the absence of counseling experience (Cluster~D) and non-participation in school self-governance and extracurricular activities (Cluster~A), for which vulnerability response rates remain elevated. The high prevalence of no counseling experience may reflect stable mental states with low demand for counseling services, though this interpretation requires verification through the school-specific analysis discussed in Section~\ref{sec:interaction_map}.

\paragraph*{School 6: Academic Stress and Emotional Distress.}
Figure~\ref{fig:u06} shows that the majority of students in School~6 exhibit negative interactions with Cluster~C. However, a few students positioned near Cluster~C display notably large vector magnitudes, indicating severe vulnerabilities in stress, depression, and smartphone dependency. These outlying students drive the School~6 vector to align closely with Cluster~C items in Figure~\ref{fig:uk_map}, despite the majority of students being positioned in the opposite direction.

Student~26 illustrates how the inner product formulation leverages both direction and magnitude. This student exhibits a notably large vector magnitude, enabling the model to capture simultaneous strong interaction with Cluster~C and negative interaction with Cluster~B. In practice, Student~26 was the only student in School~6 who reported vulnerabilities across all stress items yet showed adequate outcomes in the HP, SAD, and EM categories that characterize Cluster~B. Section~8 of the Supplementary Material further demonstrates that while Student~26 shares similar cosine similarity with Students~2 and~27, the inner product reveals substantial differences in interaction strength due to differences in vector magnitude.

The concentration of Cluster~C vulnerabilities in identifiable students supports targeted intervention design. School~6 should prioritize interventions addressing academic stress, depression, and apathy for these high-risk students. Notably, the same students are positioned opposite Cluster~A, indicating that they do not experience substantial difficulties in study planning and implementation. Interventions should therefore focus on alleviating burden related to workload or grade pressure rather than academic performance difficulties. Recommended approaches include process-oriented instruction that emphasizes learning over performance, cultivation of supportive school environments that reduce competitive pressure, and stress management programs. Additionally, to prevent smartphone addiction from developing as a stress-coping mechanism, digital health education promoting risk awareness is warranted.

\paragraph*{School 9: Heterogeneous Student Needs.}
School~9 exhibits the highest $\tilde{\alpha}_{(k)}$ in Figure~\ref{fig:tilde_alphak}, but its vulnerability profile differs qualitatively from that of School~6. Whereas School~6 exhibited outlying vulnerabilities concentrated in a specific mental health domain, School~9 exhibits a broader spectrum of vulnerabilities distributed across many students. Figure~\ref{fig:u09} shows that student vectors are generally small in magnitude, with the majority positioned opposite the main concentration of item vectors. However, approximately 25\% of students are positioned notably closer to Clusters~A, B, and~C than in other schools, suggesting that overall stability coexists with elevated vulnerability in a substantial minority.

This pattern results in the School~9 vector being oriented opposite to the majority of item vectors, yet with a relatively small magnitude compared to proximate school vectors in Figure~\ref{fig:uk_map}. The small magnitude reflects the offsetting contributions of the stable majority and the vulnerable minority, rather than uniformly moderate vulnerability. As in School~6, Student~14, positioned near Cluster~C, displays a large vector magnitude, reflecting elevated stress and smartphone dependency combined with stable competencies in the domains predominant in Cluster~B.

This heterogeneous profile warrants a differentiated intervention strategy. School~9 should implement broad-based mental health support for all students while providing targeted professional support to those exhibiting severe difficulties. Individual student interaction maps for all 35 schools are available in Section~9 of the Supplementary Material.

\section{Conclusion}\label{sec:conclusion}

This study developed the hierarchical latent space item response model (HLSIRM), a framework that extends latent space item response modeling to hierarchically structured data, and applied it to mental health vulnerability data from 2,210 elementary school students across 35 schools in Incheon, South Korea.

HLSIRM makes two methodological contributions to the analysis of hierarchical item response data. First, it introduces explicit hierarchical structure for respondent parameters, enabling decomposition of response patterns into school-level tendencies and individual-level variation. This hierarchical specification distinguishes HLSIRM from existing approaches that either estimate only individual-level parameters and aggregate post hoc, or fit separate models per group and embed them into a common space. By estimating school-level and individual-level parameters simultaneously, HLSIRM ensures that group-level summaries are optimized to represent their constituent members, while uncertainty at both levels is properly propagated. Second, the inner product formulation captures both positive and negative interactions between respondents and items, providing richer interpretation than distance-based approaches that restrict associations to positive homophily. This formulation decouples the direction and magnitude of latent vectors, allowing the model to distinguish between the type and strength of respondent-item interactions. Together, these features enable direct visualization of school-item relationships in a unified interaction map, supporting identification of school-specific vulnerability domains that inform targeted intervention design.

The application to Incheon elementary school data demonstrated the practical utility of these methodological features. The interaction-adjusted parameters revealed that the majority of schools demonstrate generally adequate mental health when aggregated across all domains, but that interaction effects substantially influence vulnerability assessment: some schools exhibit elevated vulnerability driven by alignment with specific item clusters rather than generalized difficulties. Clustering item vectors by directional similarity identified four empirically derived vulnerability domains---disengagement and non-participation, psychosocial difficulties, stress-depression and digital coping, and counseling experience---several of which cut across predefined survey categories, confirming that the inner product formulation captures empirical associations not reflected in the instrument's categorical structure. The school-specific analysis demonstrated that HLSIRM distinguishes qualitatively different vulnerability profiles---domain-concentrated vulnerabilities driven by outlying students (School~6), heterogeneous needs requiring differentiated strategies (School~9), and generally stable mental health (School~35)---each of which implies distinct intervention approaches. Associations between school-level educational programs and student positions in the interaction map provide construct validation, suggesting that the identified patterns reflect genuine institutional characteristics rather than statistical artifacts.

The analysis also revealed that mental health patterns vary systematically across Incheon's regional types, with rural island schools exhibiting response patterns distinct from urban and industrial area schools. These regional differences underscore the need for context-sensitive mental health support and suggest that educational authorities should consider local socioeconomic conditions when designing intervention programs.

Several limitations should be acknowledged. First, the cross-sectional design precludes causal inference about relationships between school characteristics and student mental health patterns. While we observed associations between educational programs and student outcomes, longitudinal designs would be necessary to establish temporal precedence and identify causal mechanisms. Second, the model preserves measurement invariance by specifying item parameters without hierarchical structure. While the empirical analysis demonstrated that item clustering transcends predefined categories in substantively meaningful ways, future extensions could explore hierarchical item structures---for instance, incorporating the predefined categorical domains as an upper level---while maintaining essential measurement invariance properties.

\section*{Acknowledgments}

This work was partially supported by the National Research Foundation of Korea [grant number NRF 2020R1A2C1A01009881, NRF-2021S1A3A2A03088949, RS-2023-00217705, RS-2024-00333701; Basic Science Research Program awarded to IHJ], [grant number NRF-2025S1A5C3A0200632411; awarded to IHJ and MJ], and the ICAN (ICT Challenge and Advanced Network of HRD) support program [grant number RS-2023-00259934], supervised by the IITP (Institute of Information \& Communications Technology Planning \& Evaluation). Correspondence should be addressed to Ick Hoon Jin, Department of Applied Statistics, Department of Statistics and Data Science, Yonsei University, Seoul, Republic of Korea. E-Mail: ijin@yonsei.ac.kr.

\bibliographystyle{Chicago}
\bibliography{reference}

\end{document}